\documentclass[twocolumn]{aastex62}

\received{May 27, 2018}
\revised{May 27, 2018}
\accepted{\today}
\submitjournal{ApJ}

\shorttitle{Occultations by the TNO 2003~VS$_2$}
\shortauthors{Benedetti-Rossi et al.}

\begin{document}

\title{The trans-Neptunian object (84922) 2003~VS$_2$ through stellar occultations}

\correspondingauthor{G. Benedetti-Rossi}
\email{gugabrossi@gmail.com, gustavo.benedetti-rossi@obspm.fr, gustavorossi@on.br}

\author[0000-0002-4106-476X]{Gustavo Benedetti-Rossi}
\affil{Observat\'orio Nacional (ON/MCTIC), Rua Gal. Jos\'e Cristino, 77 - Bairro Imperial de S\~ao Crist\'ov\~ao, Rio de Janeiro, Brazil}
\affil{Laborat\'orio Interinstitucional de e-Astronomia (LIneA) \& INCT do e-Universo, Rio de Janeiro, Brazil}
\affil{LESIA, Observatoire de Paris -- Section Meudon, 5 Place Jules Janssen -- 92195 Meudon Cedex}

\author[0000-0002-1123-983X]{P. Santos-Sanz}
\affil{Instituto de Astrof\'isica de Andaluc\'ia, IAA-CSIC, Glorieta de la Astronom\'ia s/n, 18008 Granada, Spain}

\author{J. L. Ortiz}
\affil{Instituto de Astrof\'isica de Andaluc\'ia, IAA-CSIC, Glorieta de la Astronom\'ia s/n, 18008 Granada, Spain}

\author[0000-0002-8211-0777]{M. Assafin}
\affil{Observat\'orio do Valongo/UFRJ, Ladeira Pedro Antonio 43, Rio de Janeiro, 20080-090, Brazil}

\author[0000-0003-1995-0842]{B. Sicardy}
\affil{LESIA, Observatoire de Paris -- Section Meudon, 5 Place Jules Janssen -- 92195 Meudon Cedex}
\affil{PSL Research University, CNRS, Sorbonne Universit\'e, UPMC Univ. Paris 06, Univ. Paris Diderot, Sorbonne Paris Cit\'e}

\author{N. Morales}
\affil{Instituto de Astrof\'isica de Andaluc\'ia, IAA-CSIC, Glorieta de la Astronom\'ia s/n, 18008 Granada, Spain}

\author[0000-0003-1690-5704]{R. Vieira-Martins}
\affil{Observat\'orio Nacional (ON/MCTIC), Rua Gal. Jos\'e Cristino, 77 - Bairro Imperial de S\~ao Crist\'ov\~ao, Rio de Janeiro, Brazil}
\affil{Laborat\'orio Interinstitucional de e-Astronomia (LIneA) \& INCT do e-Universo, Rio de Janeiro, Brazil}
\affil{IMCCE/Observatoire de Paris, CNRS UMR 8028, Paris, France}

\author[0000-0001-5963-5850]{R. Duffard}
\affil{Instituto de Astrof\'isica de Andaluc\'ia, IAA-CSIC, Glorieta de la Astronom\'ia s/n, 18008 Granada, Spain}

\author[0000-0003-2311-2438]{F. Braga-Ribas$^{8,1,2,3}$}
\affil{Federal University of Technology - Paran\'a (UTFPR/DAFIS), Av. Sete de Setembro, 3165, CEP 80230-901, Curitiba/PR, Brazil}
\affil{Observat\'orio Nacional (ON/MCTIC), Rua Gal. Jos\'e Cristino, 77 - Bairro Imperial de S\~ao Crist\'ov\~ao, Rio de Janeiro, Brazil}
\affil{Laborat\'orio Interinstitucional de e-Astronomia (LIneA) \& INCT do e-Universo, Rio de Janeiro, Brazil}
\affil{LESIA, Observatoire de Paris -- Section Meudon, 5 Place Jules Janssen -- 92195 Meudon Cedex}

\author[0000-0002-6085-3182]{F. L. Rommel}
\affil{Observat\'orio Nacional (ON/MCTIC), Rua Gal. Jos\'e Cristino, 77 - Bairro Imperial de S\~ao Crist\'ov\~ao, Rio de Janeiro, Brazil}
\affil{Laborat\'orio Interinstitucional de e-Astronomia (LIneA) \& INCT do e-Universo, Rio de Janeiro, Brazil}

\author[0000-0002-1642-4065]{J. I. B. Camargo}
\affil{Observat\'orio Nacional (ON/MCTIC), Rua Gal. Jos\'e Cristino, 77 - Bairro Imperial de S\~ao Crist\'ov\~ao, Rio de Janeiro, Brazil}
\affil{Laborat\'orio Interinstitucional de e-Astronomia (LIneA) \& INCT do e-Universo, Rio de Janeiro, Brazil}

\author[0000-0002-2193-8204]{J. Desmars}
\affil{LESIA, Observatoire de Paris -- Section Meudon, 5 Place Jules Janssen -- 92195 Meudon Cedex}

\author[0000-0002-0764-5042]{A. F. Colas}
\affil{IMCCE/Observatoire de Paris, CNRS UMR 8028, Paris, France}

\author{F. Vachier}
\affil{IMCCE/Observatoire de Paris, CNRS UMR 8028, Paris, France}

\author[0000-0002-5045-9675]{Alvarez-Candal}
\affil{Observat\'orio Nacional (ON/MCTIC), Rua Gal. Jos\'e Cristino, 77 - Bairro Imperial de S\~ao Crist\'ov\~ao, Rio de Janeiro, Brazil}

\author[0000-0003-2132-7769]{E. Fern\'andez-Valenzuela}
\affil{Florida Space Institute, University of Central Florida, 12354 Research Parkway, Partnership 1, Suite 211, Orlando, FL, USA}

\author{L. Almenares}
\affil{Departamento de Astronom\'ia, Facultad de Ciencias, Montevideo, Uruguay}

\author{R. Artola}
\affil{Estaci\'on Astrof\'isica de Bosque Alegre, Cordoba, Argentina}

\author{T.-P. Baum}
\affil{Observatoire Astronomique des Makes, Les Makes, 97421 -- La Rivi\`ere, Ile de la R\'eunion, France}

\author{R. Behrend}
\affil{Observatoire de Gen\`eve, CH-1290 Sauverny, Switzerland}

\author{D. B\'erard}
\affil{LESIA, Observatoire de Paris -- Section Meudon, 5 Place Jules Janssen -- 92195 Meudon Cedex}

\author{F. Bianco}
\affil{University of Delaware Department of Physics and Astronomy}
\affil{University of Delaware Joseph R. Biden Jr. School for Public Policy and Administration}
\affil{University of Delaware Data Science Institute}
\affil{New York University Center for Urban Science and Progress}

\author{N. Brosch}
\affil{School of Physics \& Astronomy and Wise Observatory, Tel Aviv University, Israel}


\author{A. Ceretta}
\affil{Observatorio Astron\'omico Los Molinos, MEC, Uruguay}

\author{C. A. Colazo}
\affil{Estaci\'on Astrof\'isica de Bosque Alegre, Cordoba, Argentina}

\author[0000-0002-3362-2127]{A. R. Gomes-Junior$^{20,2}$}
\affil{UNESP - S\~ao Paulo State University, Grupo de Din\^amica Orbital e Planetologia, CEP 12516-410, Guaratinguet\'a, SP, Brazil}
\affil{Laborat\'orio Interinstitucional de e-Astronomia (LIneA) \& INCT do e-Universo, Rio de Janeiro, Brazil}

\author{V. D. Ivanov}
\affil{ESO, Karl-Schwarzschild-Str. 2, 85748 Garching bei M\"unchen, Germany}

\author{E. Jehin}
\affil{STAR Institute - University of Li\`ege, Alle\'́e du 6 Ao\^ut 17, B-4000 Li\`ege, Belgium}

\author{S. Kaspi}
\affil{School of Physics \& Astronomy and Wise Observatory, Tel Aviv University, Israel}

\author{J. Lecacheux}
\affil{LESIA, Observatoire de Paris -- Section Meudon, 5 Place Jules Janssen -- 92195 Meudon Cedex}

\author{A. Maury}
\affil{San Pedro de Atacama Celestial Explorations, Casilla 21, San Pedro de Atacama 1410000, Chile}

\author{R. Melia}
\affil{Estaci\'on Astrof\'isica de Bosque Alegre, Cordoba, Argentina}

\author{S. Moindrot}
\affil{Observatoire de Puimichel, 04700 -- Puimichel, France}

\author[0000-0003-0088-1808]{B. Morgado}
\affil{Observat\'orio Nacional (ON/MCTIC), Rua Gal. Jos\'e Cristino, 77 - Bairro Imperial de S\~ao Crist\'ov\~ao, Rio de Janeiro, Brazil}
\affil{Laborat\'orio Interinstitucional de e-Astronomia (LIneA) \& INCT do e-Universo, Rio de Janeiro, Brazil}

\author{C. Opitom}
\affil{European Southern Observatory - Alonso de Cordova 3107, Vitacura, Santiago Chile}

\author{A. Peyrot}
\affil{Observatoire Astronomique des Makes, Les Makes, 97421 -- La Rivi\`ere, Ile de la R\'eunion, France}

\author{J. Pollock}
\affil{Physics and Astronomy Department, Appalachian State University, Boone, North Carolina 28608, USA}

\author{A. Pratt}
\affil{IOTA/ES -- International Occultation Timing Association / European Section}

\author{S. Roland}
\affil{CURE, Universidad de la Republica, Uruguay}

\author{J. Spagnotto}
\affil{Observatorio El Catalejo, Santa Rosa, La Pampa, Argentina}

\author{G. Tancredi}
\affil{Departamento de Astronom\'ia, Facultad de Ciencias, Montevideo, Uruguay}

\author{J.-P. Teng}
\affil{Observatoire Astronomique des Makes, Les Makes, 97421 -- La Rivi\`ere, Ile de la R\'eunion, France}

\author{P. Cacella}
\affil{DogsHeaven Observatory X87, Bras\'ilia, Brazil}

\author{M. Emilio}
\affil{Universidade Estadual de Ponta Grossa, O.A. - DEGEO, Avenida Carlos Cavalcanti 4748, Ponta Grossa 84030-900, PR, Brazil}
\affil{Observat\'orio Nacional (ON/MCTIC), Rua Gal. Jos\'e Cristino, 77 - Bairro Imperial de S\~ao Crist\'ov\~ao, Rio de Janeiro, Brazil}

\author{F. Feys}
\affil{IOTA/ES -- International Occultation Timing Association / European Section}

\author{R. Gil-Hutton}
\affil{Grupo de Ciencias Planetarias, Departamento de Geof\'isica y Astronom\'ia, San Juan National University and CONICET, Av. Jos\'e I. de la Roza 590(0), J5402DCS, San Juan, Argentina}

\author{C. Jacques}
\affil{Southern Observatory for Near Earth Asteroids Research (SONEAR)}

\author{D. I. Machado}
\affil{ Universidade Estadual do Oeste do Paran\'a (Unioeste), Avenida Tarqu\'inio Joslin dos Santos 1300, Foz do Iguaçu, PR  85870-650, Brazil}
\affil{Polo Astron\^omico Casimiro Montenegro Filho/FPTI-BR, Avenida Tancredo Neves 6731, Foz do Iguaçu, PR 85867-900, Brazil}

\author{M. Malacarne}
\affil{Universidade Federal do Esp\'irito Santo (UFES)}

\author{I. Manulis}
\affil{Weizmann Institute of Science, Rehovot, Israel}

\author{A. C. Milone}
\affil{ Astrophysics Division, National Institute for Space Research (INPE), Av. dos Astronautas 1758, S\~ao Jos\'e dos Campos, SP, 12227-010, Brazil}

\author{G. Rojas}
\affil{Observat\'orio Astron\^omico - Universidade Federal de S\~ao Carlos (UFSCar), S\~ao Carlos, Brazil}

\author[0000-0002-4939-013X]{R. Sfair}
\affil{UNESP - S\~ao Paulo State University, Grupo de Din\^amica Orbital e Planetologia, CEP 12516-410, Guaratinguet\'a, SP, Brazil}


\begin{abstract}

We present results from three world-wide campaigns that resulted in the detections of two single-chord and one multi-chord stellar occultations by the Plutino object (84922) 2003~VS$_2$. From the single-chord occultations in 2013 and 2014 we obtained accurate astrometric positions for the object, while from the multi-chord occultation on November 7th, 2014, we obtained the parameters of the best-fitting ellipse to the limb of the body at the time of occultation. We also obtained short-term photometry data for the body in order to derive its rotational phase during the occultation. The rotational light curve present a peak-to-peak amplitude of 0.141 $\pm$ 0.009 mag. This allows us to reconstruct the three-dimensional shape of the body, with principal semi-axes
$a = 313.8 \pm 7.1$ km, 
$b = 265.5^{+8.8}_{-9.8}$ km, and
$c = 247.3^{+26.6}_{-43.6}$ km, which is not consistent with a Jacobi triaxial equilibrium figure. The derived spherical volume equivalent diameter of
$548.3 ^{+29.5}_{-44.6}$ km
is about 5\% larger than the radiometric  diameter of 2003~VS$_2$ derived from Herschel data of $523 \pm 35$ km, but still compatible with it within error bars. From those results we can also derive the geometric albedo ($0.123 ^{+0.015}_{-0.014}$) and, under the assumption that the object is a Maclaurin spheroid, the density $\rho = 1400^{+1000}_{-300}$ for the plutino. The disappearances and reappearances of the star during the occultations do not show any compelling evidence for a global atmosphere considering a pressure upper limit of about 1 microbar for a
pure nitrogen atmosphere, nor secondary features (e.g. rings or satellite)  around the main body.

\end{abstract}

\keywords{Trans-Neptunian Object -- Kuiper Belt -- (84922) 2003~VS$_2$ -- Rotation Light Curve -- Stellar Occultation}


\section{Introduction \label{sec:Introduction}}

Trans-Neptunian Objects (TNOs) are bodies that orbit the Sun with orbital semi-major axis larger than that of Neptune \citep{Jewitt2008}. Due to their large distance to the Sun and low spatial density, those objects do not experience extensive differentiation.
Consequently, TNOs are the least evolved bodies in the Solar System, at least from the composition point of view. Hence, knowing their physical parameters such as size, shape, albedo, density, presence of atmosphere, rings, and their evolution, yield important information on the nature of material, physical conditions, and history of the primitive solar nebula \citep{Morbidelli2008}, and about the formation and evolution of our solar system \citep{LykawkaMukai2008, Parker2015, Santos-Sanz2016}. Additionally, the Kuiper Belt provides the natural connection with the study of protoplanetary disks observed around other stars \citep{Anglada2017}.

Despite the fact that more than 25 years have elapsed since this population was discovered (the first TNO discovered was (15760) Albion by \cite{Jewitt1993}, only Pluto and, more recently (in January 1st, 2019), the small TNO 2014~MU$_{69}$ have been visited so far by a spacecraft, the NASA/New Horizons mission \citep{Stern2019}, making them unique bodies in the trans-Neptunian region with a more complete set of information. In summary, our knowledge of basic physical properties of the TNO population is still scarce and fragmentary, mainly due to the faintness and small angular sizes of these bodies as seen from Earth \citep{Stansberry2008, Lellouch2013}.

In this context, we have been conducting various kinds of Earth-based observations for nearly two decades to gather relevant physical information about TNOs \citep{Lellouch2002,Ortiz2002,Ortiz2004,Ortiz2006,Ortiz2007,Ortiz2011,Ortiz2012,Ortiz2015,Ortiz2017,Belskaya2006,Duffard2008,Assafin2010,Thirouin2010,Sicardy2011,Assafin2012,Braga-Ribas2013,Braga-Ribas2014,Camargo2014,Benedetti-Rossi2016,Berard2017,Dias-Oliveira2017,Leiva2017,Camargo2018}. In this framework, we predicted and detected three stellar occultations by the TNO (84922) 2003~VS$_2$ -- VS2 hereafter -- that we monitored through a  collaboration with a large panel of amateurs. Among the three events, we obtained two single-chord detections in December 2013 and in March 2014 and one multi-chord detection in November 2014. The latter observation was made from four well separated sites, from which valuable physical information is derived, as described in this paper.

The object VS2 was discovered by the Near Earth Asteroid Tracking (NEAT)\footnote{\url{https://neat.jpl.nasa.gov/}} program on November 14, 2003. As a 2:3 resonant object with Neptune, it  belongs to the plutino class \citep{Gladman2008, MPEC2006X45}\footnote{Another classification is given by M. Brown in \url{http://web.gps.caltech.edu/~mbrown/dps.html}}. Most of its physical properties were derived from thermal measurements using the space telescopes Herschel and Spitzer \citep{Stansberry2008, Mommert2012,Santos-Sanz2017}. Orbital parameters, absolute magnitude, B--R color, rotation period, photometric variation, taxonomy, diameter, and geometric albedo, taken from previously published works, are listed in Table~\ref{tab:previous_publications}.

\begin{deluxetable*}{cccccccccccc}
\tablecaption{Orbital and physical characteristics of the plutino object 2003~VS$_2$.\label{tab:previous_publications}}
\tablewidth{700pt}
\tabletypesize{\scriptsize}
\tablehead{
\colhead{a} & \colhead{q} & \colhead{i} & \colhead{e} & \colhead{H$_{v}$*} & \colhead{B -- R} & \colhead{P} & \colhead{$\Delta$m} & \colhead{Taxon.} & \colhead{D**} & \colhead{p$_{V}$} & \colhead{APmag} \\
$[$AU$]$ & $[$AU$]$ & $[$deg$]$ &  & $[$mag$]$ & $[$mag$]$ & $[$h$]$ & $[$mag$]$  &  & $[$km$]$ & $[$\%$]$
}
\startdata
39.25 & 36.43 & 14.80 & 0.07 & 4.130 $\pm$ 0.070 & 1.52 $\pm$ 0.03 & 7.4175285 $\pm$ 0.00001 & 0.224 $\pm$ 0.013 & BB & 523.0$^{+35.1}_{-34.4}$ & 14.7$^{+6.3}_{-4.3}$ & 19.95
\enddata
\tablecomments{
Orbital parameters: \textbf{a}: semimajor axis in astronomical units (au); \textbf{q}: perihelion distance in au; \textbf{i}: orbital inclination in degrees; \textbf{e}: eccentricity -- from ``JPL Small-Body Database Browser''. \textbf{H$_{v}$}: average absolute magnitude at V-band obtained from \cite{AlvarezCandal2016} and priv. comm.; \textbf{B -- R} color from \cite{Sheppard2007, Mommert2012} and references therein. \textbf{P}: preferred rotation period in hours from \cite{Santos-Sanz2017}. \textbf{$\Delta$m $[mag]$}: light curve peak-to-valley amplitude from \cite{Thirouin2013}. \textbf{Taxon.}: taxonomic color class from \cite{Perna2010}, and references therein. \textbf{D}: area-equivalent diameter and \textbf{p$_{V}$}: geometric albedo at V-band from \cite{Mommert2012} using Herschel and Spitzer data. \textbf{APmag}: 2003~VS$_2$'s average apparent visual magnitude between December 2013 and December 2014 from Horizons/JPL (calculated using H$_{v}$ = 4.2). * \cite{Sheppard2007, Mommert2012} obtained a value of H$_{v}$ = 4.110 $\pm$ 0.380; ** \cite{Stansberry2008} obtained a diameter D = 725 $^{+199}_{-187}$ km with Spitzer data only.}
\end{deluxetable*}

Here we present the results from those three stellar occultations by VS2, with the determination of size and shape of this body based on the multi-chord occultation of November 07, 2014. We also present a rotational light curve for VS2 derived from the 1.5~m telescope located at Sierra Nevada observatory in Granada (Spain), showing an amplitude that is smaller than reported previously in the literature.

We describe in Section~\ref{sec:observation} the circumstances of observation for the stellar occultations (predictions, updates and the occultations themselves) and for the photometric runs to obtain the light curve. In Section~\ref{sec:analysis} we describe our data reduction, analysis and the results obtained from the stellar occultations. We discuss in Section~\ref{sec:results} the results concerning the 3D-shape of the body as well as other features by combining our occultation and photometric data. Finally, conclusions are presented in Section~\ref{sec:conclusions}.

\section{Observations} \label{sec:observation}

Observations performed in this work are separated into three groups: (\ref{subsec:occultation_prediction}) observations performed to predict and refine the the astrometric positions of the occulted stars and of the object for the occultations;  (\ref{subsec:occultation}) observations of the  occultations; and (\ref{subsec:photometric_observation}) the observation runs that yielded VS2's rotational light curve.

\subsection{Stellar Occultations Predictions and Updates} \label{subsec:occultation_prediction}

Stellar occultations require large efforts in order to determine accurate positions of the star and of the object's ephemeris. Despite the fact that the \textit{Gaia} catalog -- on its second data release, GDR2 -- presents positions and proper motions for stars with unprecedented accuracy (milliarcsec, or mas, levels \citep{GAIA2016a,GAIA2016,GAIA2018,vanLeeuwen2017}), the object's ephemeris is usually determined at accuracies of a few hundred mas, due to scarce observations. This is larger than the projected sizes of the objects in the sky plane, resulting in misses when attempting stellar occultation obervations. 
Until data from large surveys, such as the LSST\footnote{\url{https://www.lsst.org/}}, become available in a few years from now, regular observations of those small objects will still be needed for most of them in order to improve their orbital parameters. In fact, even with LSST data available, big efforts will still be needed to improve objects' ephemeris, as LSST will mostly cover the southern sky, leaving the objects in the northern hemisphere with big uncertainties. Note also that all those objects have long orbital periods (nearly two centuries for the closest TNOs) and that the time-span of the observations are short (less than 20 years) so the uncertainties in the orbital elements are very sensitive to new observations. 

At the time of the three occultations reported here, we did not have the \textit{Gaia} catalog in our hands so the candidate stars were identified in systematic surveys performed at the 2.2~m telescope of the European Southern Observatory (ESO) at La Silla -- IAU code 809 --, using the Wide Field Imager (WFI). The surveys yielded local astrometric catalogues for 5 Centaurs and 34 TNOs (plus Pluto and its moons) up to 2015, with stars with magnitudes as faint as R mag $\sim$19, see details in \cite{Assafin2010,Assafin2012} and \cite{Camargo2014}. 
All three VS2 occultation candidate stars were then identified in the GDR2 catalog and their right ascension, declination, proper motions, and G, B, V and K magnitudes are presented in Table~\ref{tab:occ_stars_RA_DEC_Gmag}.

\begin{deluxetable*}{cccccccccccc}
\tablecaption{J2000 stars coordinates, proper motions, parallax, and magnitudes \label{tab:occ_stars_RA_DEC_Gmag}}
\tablewidth{700pt}
\tabletypesize{\scriptsize}
\tablehead{
\colhead{Occultation Date} & \colhead{RA ICRS} & \colhead{errRA} & \colhead{DEC ICRS} & \colhead{errDEC} & \colhead{pmRA} & \colhead{pmDEC} & \colhead{Plx} & \colhead{G} & \colhead{B} & \colhead{V} & \colhead{K} \\
 &  & (mas) &  & (mas) & (mas/yr) & (mas/yr) & (mas) & (mag) & (mag) &(mag) & (mag)
}
\startdata
December 12, 2013 & 04$^h$ 36$^m$ 19$^s$.830879 & 0.0502 & +33$^\circ$ 55' 55.45028 & 0.0434 & 0.878 & -0.649 & 0.0733 & 16.0899 & 16.57 & 16.20 & 12.64 \\
%
March 04, 2014 & 04$^h$ 32$^m$ 04$^s$.463481 & 0.0548 & +33$^\circ$ 24' 41.95395 & 0.0422 & 2.925 & -3.268 & 0.7682 & 15.9110 & 17.29 & 16.23 & 13.54 \\
%
November 07, 2014 & 04$^h$ 48$^m$ 32$^s$.138835 & 0.0406 & +33$^\circ$ 58' 36.40859 & 0.0303 & 7.404 & -8.135 & 1.7609 & 15.1311 & 17.16 & 15.82 & 12.04 \\
%
\enddata
\tablecomments{
\textbf{RA} and \textbf{DEC} ICRS: Barycentric right ascension and declination (ICRS) propagated to the occultation epoch.; \textbf{errRA} and \textbf{errDEC}: Standard error of right ascension (multiplied by cosDEC) and declination; \textbf{pmRA} and \textbf{pmDEC}: proper motion in right ascension (multiplied by cosDEC) and in declination direction; \textbf{Plx}: Absolute stellar parallax; G magnitude. All values obtained from GDR2 \citep{GAIA2016a,GAIA2016,GAIA2018}. B, V, and K magnitudes obtained from NOMAD catalog \citep{Zacharias2004}.}
\end{deluxetable*}

Astrometric updates of the candidate stars (and the TNO when possible) were performed using several telescopes close to the dates of events in Brazil at Pico dos Dias Observatory -- IAU code 874 -- (Perkin Elmer 1.6~m, Boller \& Chivens 0.6~m, and ZEISS 0.6~m), in Spain at Calar Alto Observatory -- IAU code 493 -- (1.23~m and 2.2~m telescopes), at Sierra Nevada Observatory -- IAU code J86 -- (1.5~m telescope), and at Observatorio de La Hitta -- IAU code I95 -- (77~cm telescope), and in France at Pic du Midi -- IAU code 586 -- (T100~cm telescope) so to reduce systematic errors caused by the reference catalogs. Astrometric positions obtained for VS2 with those observations are listed on Table~\ref{tab:astrometric_pos_VS2}. The final positions obtained for the star and the TNO before the occultations provided predictions with uncertainties as large as 50 mas, equivalent to about 1300 km when projected onto Earth's surface.

\begin{deluxetable*}{cccccccccc}
\tablecaption{Astrometric positions for VS2. \label{tab:astrometric_pos_VS2}}
\tabletypesize{\scriptsize}
\tablehead{
\multicolumn{3}{c}{Date} & \multicolumn{3}{c}{RA ICRS} & \multicolumn{3}{c}{DEC ICRS} & \colhead{Site}\\
Year & Month & Day & Deg & Min & Sec & Deg & Min & Sec & $[$IAU Code$]$ }
\startdata
2014 & 02 & 01.868243 & 68 & 03 & 09.4710 & 33 & 36 & 12.4600 & J86 \\
2014 & 02 & 01.882180 & 68 & 03 & 08.9115 & 33 & 36 & 12.1370 & J86 \\
2014 & 02 & 01.886832 & 68 & 03 & 08.8125 & 33 & 36 & 11.9780 & J86 \\
2014 & 02 & 01.891829 & 68 & 03 & 08.5650 & 33 & 36 & 11.8790 & J86 \\
2014 & 02 & 01.905770 & 68 & 03 & 08.1405 & 33 & 36 & 11.4950 & J86 \\
2014 & 02 & 03.000210 & 68 & 02 & 31.7055 & 33 & 35 & 44.4050 & J86 \\
2014 & 02 & 03.004858 & 68 & 02 & 31.5690 & 33 & 35 & 44.2810 & J86 \\
2014 & 02 & 03.009507 & 68 & 02 & 31.4220 & 33 & 35 & 44.1640 & J86 \\
2014 & 02 & 03.014154 & 68 & 02 & 31.2825 & 33 & 35 & 44.0510 & J86 \\
2014 & 02 & 03.018799 & 68 & 20 & 31.0935 & 33 & 35 & 43.9310 & J86 \\
2014 & 02 & 03.023446 & 68 & 02 & 30.9375 & 33 & 35 & 43.7860 & J86 \\
2014 & 02 & 03.028092 & 68 & 02 & 30.8565 & 33 & 35 & 43.7010 & J86 \\
2014 & 02 & 03.032739 & 68 & 02 & 30.6495 & 33 & 35 & 43.5740 & J86 \\
2014 & 02 & 03.037384 & 68 & 02 & 30.5475 & 33 & 35 & 43.4220 & J86 \\
2014 & 02 & 03.042029 & 68 & 02 & 30.4290 & 33 & 35 & 43.3400 & J86 \\
2014 & 02 & 19.920970 & 67 & 58 & 05.7000 & 33 & 29 & 06.1000 & 586 \\
2014 & 02 & 19.920971 & 67 & 58 & 05.8260 & 33 & 29 & 06.2120 & 586 \\
2014 & 02 & 19.924510 & 67 & 58 & 05.7000 & 33 & 29 & 06.0000 & 586 \\
2014 & 02 & 19.924510 & 67 & 58 & 05.7705 & 33 & 29 & 06.1050 & 586 \\
2014 & 02 & 19.928070 & 67 & 58 & 05.7000 & 33 & 29 & 06.0000 & 586 \\
2014 & 02 & 19.928074 & 67 & 58 & 05.8995 & 33 & 29 & 06.0400 & 586 \\
2014 & 02 & 19.931640 & 67 & 58 & 05.7000 & 33 & 29 & 05.8000 & 586 \\
2014 & 02 & 19.931640 & 67 & 58 & 05.8170 & 33 & 29 & 05.9790 & 586 \\
2014 & 02 & 19.935207 & 67 & 58 & 05.8155 & 33 & 29 & 05.9290 & 586 \\
2014 & 02 & 19.935210 & 67 & 58 & 05.7000 & 33 & 29 & 05.8000 & 586 \\
2014 & 02 & 19.938776 & 67 & 58 & 05.8560 & 33 & 29 & 05.8680 & 586 \\
2014 & 02 & 19.938780 & 67 & 58 & 05.7000 & 33 & 29 & 05.7000 & 586 \\
2014 & 02 & 19.942340 & 67 & 58 & 05.7000 & 33 & 29 & 05.7000 & 586 \\
2014 & 02 & 19.942344 & 67 & 58 & 05.8230 & 33 & 29 & 05.7750 & 586 \\
2014 & 02 & 19.945910 & 67 & 58 & 05.8500 & 33 & 29 & 05.6000 & 586 \\
2014 & 02 & 19.945910 & 67 & 58 & 05.8620 & 33 & 29 & 05.7090 & 586 \\
2014 & 02 & 19.949476 & 67 & 58 & 05.8080 & 33 & 29 & 05.5710 & 586 \\
2014 & 02 & 19.949480 & 67 & 58 & 05.7000 & 33 & 29 & 05.4000 & 586 \\
2014 & 02 & 19.953040 & 67 & 58 & 05.8500 & 33 & 29 & 05.4000 & 586 \\
2014 & 02 & 19.953041 & 67 & 58 & 05.9040 & 33 & 29 & 05.4890 & 586 \\
2014 & 02 & 19.956608 & 67 & 58 & 05.8545 & 33 & 29 & 05.3620 & 586 \\
2014 & 02 & 19.956610 & 67 & 58 & 05.7000 & 33 & 29 & 05.2000 & 586 \\
2014 & 02 & 19.963670 & 67 & 58 & 05.8500 & 33 & 29 & 05.1000 & 586 \\
2014 & 02 & 19.967230 & 67 & 58 & 05.7000 & 33 & 29 & 05.0000 & 586 \\
2014 & 02 & 19.970800 & 67 & 58 & 05.7000 & 33 & 29 & 05.1000 & 586 \\
2014 & 02 & 23.832322 & 67 & 58 & 25.3935 & 33 & 27 & 42.0770 & I95 \\
2014 & 02 & 23.833071 & 67 & 58 & 24.9570 & 33 & 27 & 41.8390 & I95 \\
2014 & 02 & 23.835317 & 67 & 58 & 25.4850 & 33 & 27 & 41.7380 & I95 \\
2014 & 02 & 23.836792 & 67 & 58 & 24.8685 & 33 & 27 & 41.7520 & I95 \\
2014 & 02 & 23.840530 & 67 & 58 & 25.2300 & 33 & 27 & 41.8910 & I95 \\
2014 & 02 & 23.841275 & 67 & 58 & 25.2330 & 33 & 27 & 42.0330 & I95 \\
2014 & 09 & 24.148195 & 72 & 47 & 13.9545 & 33 & 53 & 41.9090 & 493 \\
2014 & 09 & 24.157750 & 72 & 47 & 13.8450 & 33 & 53 & 42.0140 & 493 \\
2014 & 09 & 24.167348 & 72 & 47 & 13.6620 & 33 & 53 & 42.1710 & 493 \\
2014 & 09 & 24.172112 & 72 & 47 & 13.5675 & 33 & 53 & 42.2450 & 493 \\
2014 & 09 & 24.181729 & 72 & 47 & 13.4475 & 33 & 53 & 42.3420 & 493 \\
2014 & 09 & 24.186538 & 72 & 47 & 13.3410 & 33 & 53 & 42.4430 & 493 \\
2014 & 09 & 24.196181 & 72 & 47 & 13.1865 & 33 & 53 & 42.5840 & 493 \\
2014 & 09 & 24.200971 & 72 & 47 & 13.1190 & 33 & 53 & 42.6250 & 493 \\
2014 & 09 & 24.205804 & 72 & 47 & 13.0065 & 33 & 53 & 42.7340 & 493 \\
2014 & 09 & 24.210612 & 72 & 47 & 12.9735 & 33 & 53 & 42.7860 & 493 \\
2014 & 09 & 24.215402 & 72 & 47 & 12.9045 & 33 & 53 & 42.8490 & 493 \\
\enddata
\tablecomments{
Positions obtained with the ESO 2p2 telescope -- IAU code 809 -- under mission 090.C-0118(A) are published in \cite{Camargo2014}.}
\end{deluxetable*}

\subsection{Stellar Occultations} \label{subsec:occultation}

For each of the three stellar occultations, an alert was triggered at several potential sites, resulting in data collected with a large diversity of instruments, see Tables \ref{tab:occultation_circumstances}, \ref{tab:occultation_circumstances_multichord}, and \ref{tab:occultation_circumstances_multichord_negative}. All sites used robust clock synchronizations, either by a Global Positioning System (GPS) or by setting up the clocks using Internet servers -- Network Time Protocol (NTP) --, and acquisition times of each image was inserted on image header or printed on individual video frames. Using any of those synchronization methods, the times should not present absolute errors larger than 1 ms \citep{DeethsBrunette2001}. At all sites with favourable weather conditions, data were collected from about ten minutes prior to the predicted occultation times until about ten minutes after those times. No filters were used in any of the observations to maximize photon fluxes, and thus signal-to-noise-ratio (SNR).

\begin{figure}[h]
\includegraphics[width=\columnwidth]{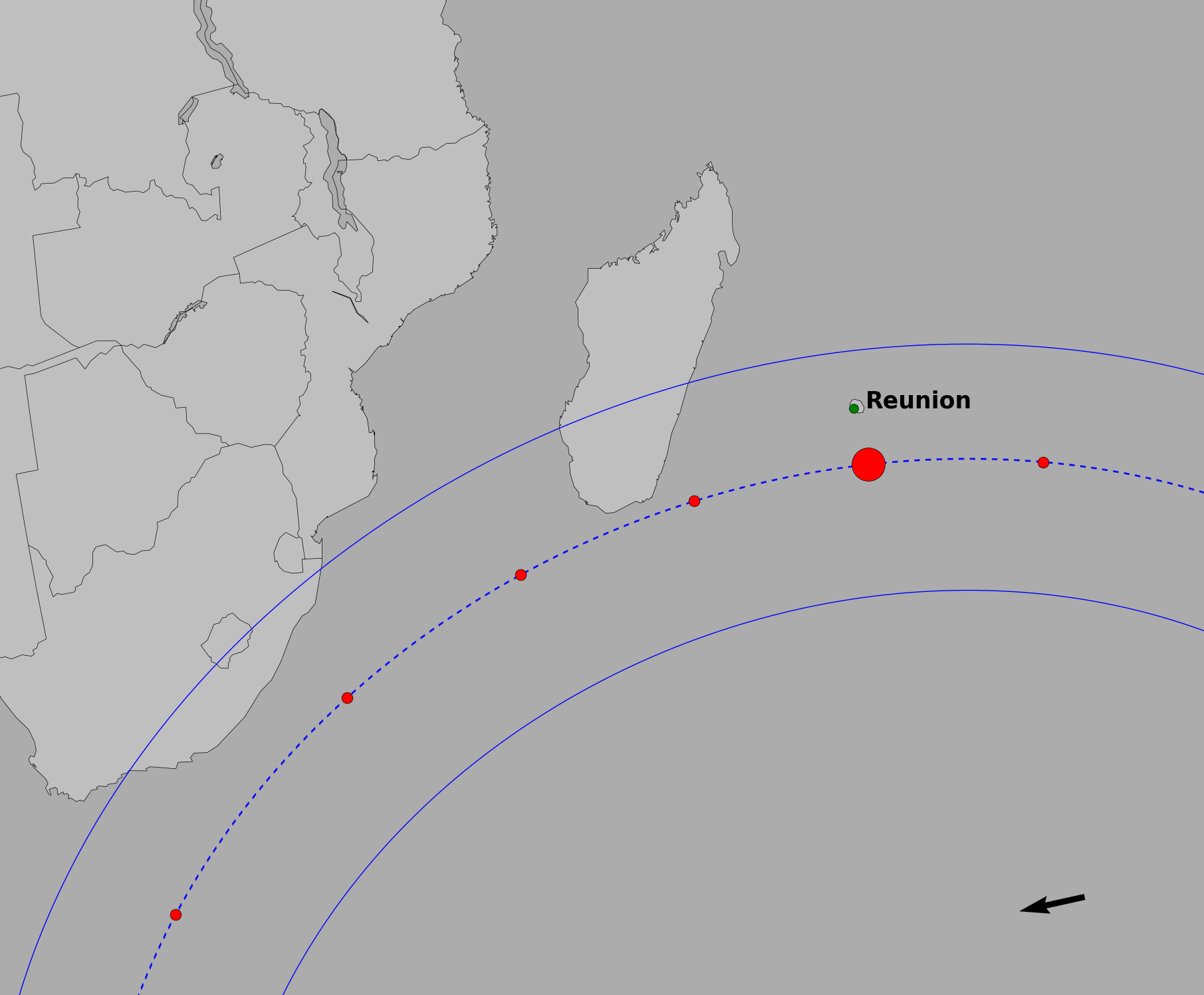}
\caption{Post-occultation map for the December 12, 2013 occultation. North is up and East is right. Blue lines represent the object equivalent diameter from the November 07, 2014 occultation of 564.8 km. Red dots represent the position of the center of the body spaced every 30 seconds, the bigger dot corresponding to 20:10:44 UTC. Direction of the shadow is shown by the arrow at the right corner. Green dot is the site position with positive detection (see Table~\ref{tab:occultation_circumstances}).
\label{fig:post_occultation_map_2013}}
\end{figure}

\begin{figure}[h]
\includegraphics[width=\columnwidth]{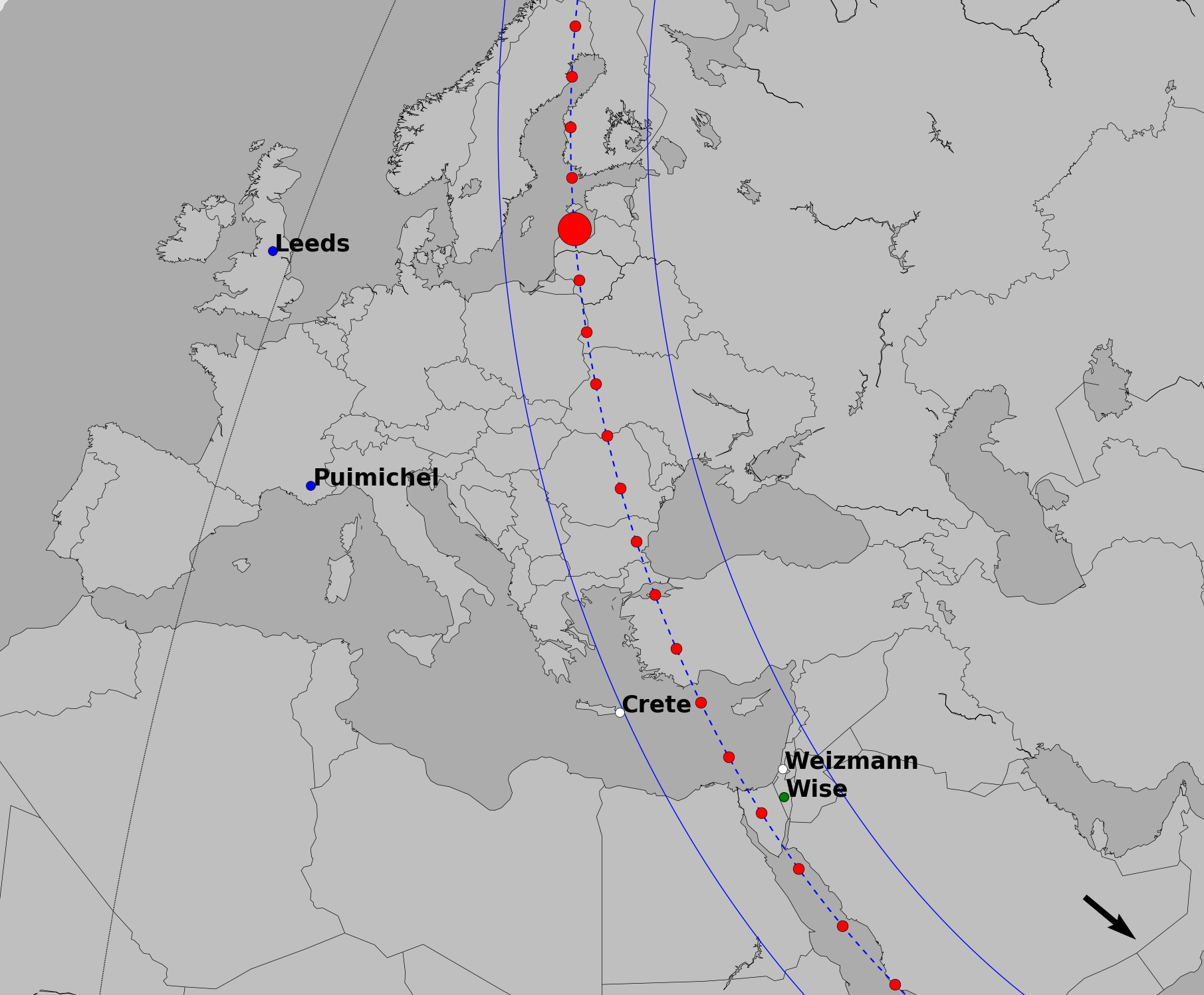}
\caption{Post-occultation map for the March 04, 2014 occultation. North is up and East is right. Blue lines represent the object equivalent diameter from the November 07, 2014 occultation of 564.8 km. Red dots represent the position of the center of the body spaced every 30 seconds, the bigger dot corresponding to 19:55:54 UTC. Direction of the shadow is shown by the arrow at the right corner. Green dot is the site position with positive detection, while the blue dots are the negative detections and the white dots are the sites with weather overcast (see Table~\ref{tab:occultation_circumstances}).
\label{fig:post_occultation_map_201403}}
\end{figure}

\begin{figure}
\includegraphics[width=\columnwidth]{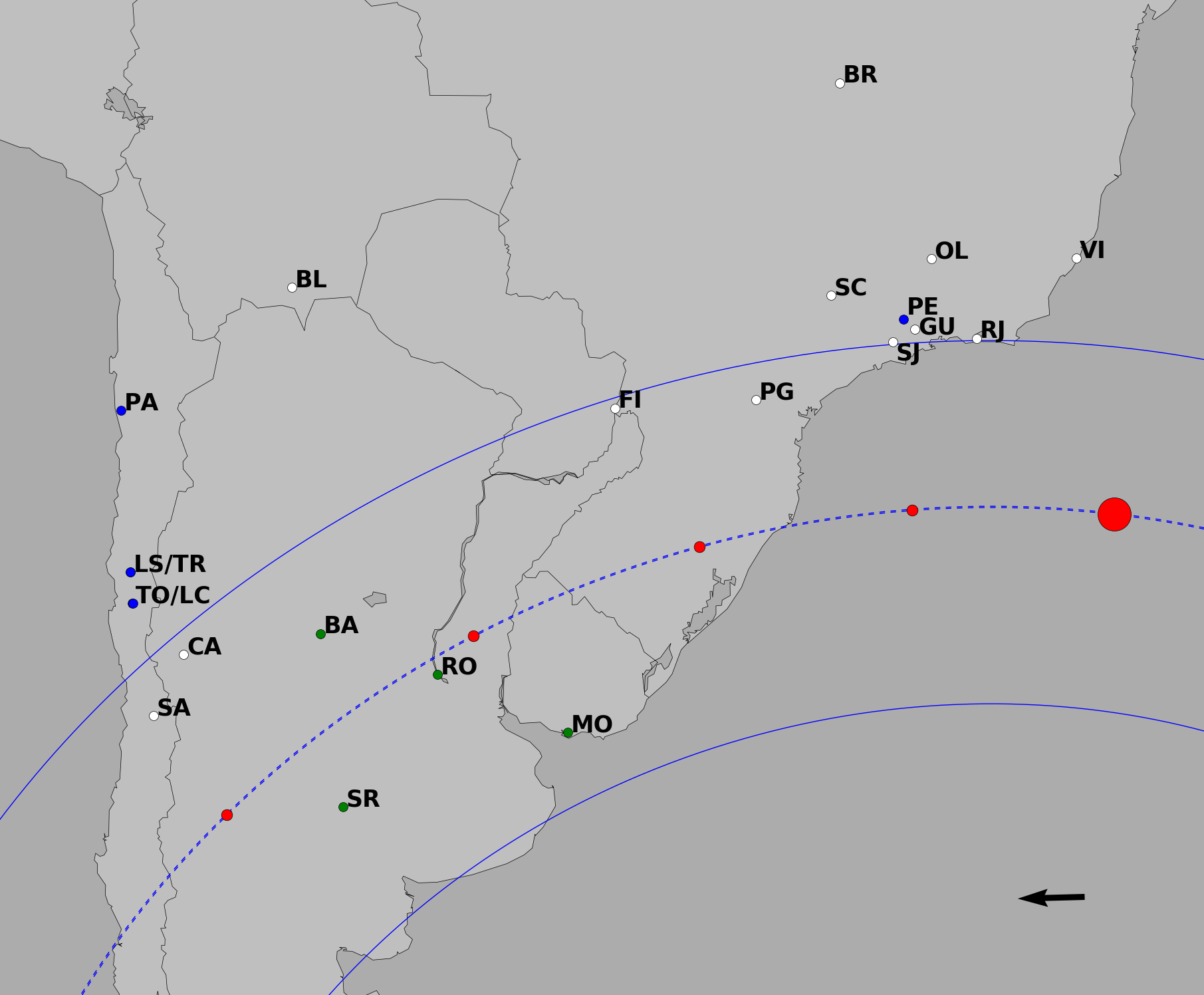}
\caption{Post-occultation map for the November 07, 2014 occultation. North is up and East is right. Blue lines represent the object equivalent diameter obtained from this occultation, of 564.8 km. Red dots represent the position of the center of the body spaced every 30 seconds with the bigger one corresponding to 04:22:00 UTC. Direction of the shadow is shown by the arrow at the right corner. Green dots are the site positions with positive detections, while the blue dots are the negative detections and the white dots are the sites with weather overcast (see Tables \ref{tab:occultation_circumstances_multichord} and \ref{tab:occultation_circumstances_multichord_negative}).
\label{fig:post_occultation_map_201411}}
\end{figure}

Out of 23 stations distributed in nine countries (France, Israel, United Kingdom, Greece, Argentina, Uruguay, Chile, Brazil and Bolivia), we obtained two single-chord events and one multi-chord occultation, for a total of 7 positive detections. Note that the March 2014 occultation was detected by two telescopes at the same site in Israel. As such, they do not provide any constraint on the shape of the object, and we consider it as a single-chord detection. Figures \ref{fig:post_occultation_map_2013}, \ref{fig:post_occultation_map_201403}, and \ref{fig:post_occultation_map_201411} show the post-occultation, reconstructed maps for the three events. Note that in those figures we used a radius for VS2 of 564.8 km (determined from the multi-chord occultation), but we do not know the rotational phase for the two single-chord occultations, so the shadow path may be smaller than this value. Details of data analysis for each occultation are given in Section \ref{sec:analysis}.

\begin{deluxetable*}{cccccc}
\tablecaption{Observation details of the single-chord stellar occultations \label{tab:occultation_circumstances}}
\tablewidth{700pt}
\tabletypesize{\scriptsize}
\tablehead{
\\
\multicolumn{6}{c}{December 12, 2013} \\
\\ \hline
Site Name & Longitude (E) & Telescope Aperture & Exposure Time &  &  \\
(Location) & Latitude (N) & Detector/Instrument & Cycle Time & Observers & Detection  \\
{$[$IAU Code$]$} & Altitude (m) &  & (s) &  & 
}
\startdata
Les Makes Observatory & 55$^\circ$ 24' 36.0'' & 60 cm & 0.75000 & J. Lecacheux &  \\
(La R\'eunion, France) & -21$^\circ$ 11' 58.4'' & Raptor Merlin 246 & 0.75031 & A. Peyrot, T.-P. Baum & Positive \\
$[181]$ & 992 &  &  & J.-P. Teng & \\
\\
\\
\\
\hline\hline
\\
\multicolumn{6}{c}{March 04, 2014}  \\
 \\ \hline
Site Name & Longitude (E) & Telescope Aperture & Exposure Time & & \\
(Location) & Latitude (N) & Detector/Instrument & Cycle Time & Observers & Detection  \\
$[$IAU Code$]$ & Altitude (m) &  & (s) &  &  \\
\hline
Wise Observatory & 34$^\circ$ 45' 43.60''   & 100 cm & 4 &  &  \\
(Mitzpe Ramon, Israel) & 30$^\circ$ 35' 50.38'' & Roper PVCAM & 7 & S. Kaspi & Positive \\
$[097]$ & 884 & PI 1300B/LN &  &  &  \\
\hline
Wise Observatory & 34$^\circ$ 45' 43.92''   & 70 cm & 5 &  &  \\
(Mitzpe Ramon, Israel) & 30$^\circ$ 35' 48.23'' & FLI & 6.5 & N. Brosch & Positive \\
$[097]$ & 865 & PL16801 &  &  &  \\
\hline
West Park Observatory & 358$^\circ$ 23' 31.98'' & 20 cm & 5.08 &  &  \\
(Leeds, UK) & 53$^\circ$ 50' 15.42''  & WATEC & 5.08 & A. Pratt & Negative \\
$[Z92]$ & 114 & WAT-910HX/RC (x256) & (video) &  & \\
\hline
Puimichel Observatory & 06$^\circ$ 01' 15.5''   & 104 cm & 0.25000 & J. Lecacheux &  \\
(Puimichel, France) & 43$^\circ$ 58' 48.7'' & Raptor Merlin 246 & 0.25031 & S. Moindrot & Negative \\
-- & 724 &  &  &  & \\
\hline
Weizmann & 34$^\circ$ 48' 45.76''   & 41 cm & -- &  &  \\
(Kraar, Israel) & 31$^\circ$ 54' 29.1'' & SBIG & -- & I. Manulis & Overcast \\
-- & 107 & ST-8XME CCD &  &  & \\
\hline
Sasteria Observatory & 25$^\circ$ 58' 16.37''   & 25 cm & -- &  &  \\
(Crete, Greece) & 35$^\circ$ 04' 11.12'' & N/A & -- & F. Feys & Overcast \\
-- & 395 &  &  &  &
\enddata
\tablecomments{
``FLI'' stands for Finger Lakes Instruments; ``PI'' for Princeton Instruments; ``PL'' for Pro Line; ``N/A'' for Not Available.}
\end{deluxetable*}

\begin{deluxetable*}{cccccc}
\tablecaption{Observation details of the multi-chord stellar occultation on November 07, 2014 with data acquisition \label{tab:occultation_circumstances_multichord}}
\tablewidth{700pt}
\tabletypesize{\scriptsize}
\tablehead{
\colhead{Site Name} & \colhead{Longitude (E)} & \colhead{Telescope Aperture} & \colhead{Exposure Time} &  &  \\
(Location) & Latitude (N) & Detector/Instrument & Cycle Time & Observers & Detection \\
$[$IAU Code$]$ & Altitude (m) &  & (s) &  & 
}
\startdata
Bosque Alegre & 295$^\circ$ 27' 01.5'' & 154 cm & 5 & C. A. Colazo & \\
(Cordoba, Argentina) & -31$^\circ$ 35' 54.0'' & CMOS ZWO ASI120MM & 5.000032 & R. Artola & Positive\\
$[821]$ & 1213.1 & & & R. Melia & \\
\hline
Ros\'ario & 299$^\circ$ 21' 09.0'' & 49 cm & 3 & & \\
(Ros\'ario, Argentina) & -32$^\circ$ 58' 16'' & CCD KAF 8300 & 4 & V. Buso & Positive \\
- & 31.0 &  &  & & \\
\hline
Montevideo & 303$^\circ$ 48' 37.1 & 46 cm & 3 & G. Tancredi &  \\
(Montevideo, Uruguay) & -34$^\circ$ 45' 20.0'' & FLI & 4 & S. Roland & Positive \\
- & 130.0 & PL9000 &  & L. Almenares & \\
 & & & & A. Ceretta & \\
\hline
Santa Rosa & 295$^\circ$ 40' 32.5'' & 20 cm & 25 & & \\
(Santa Rosa, Argentina) & -36$^\circ$ 38' 16.0''  & CCD Meade DSI-I  & 26 & J. Spagnotto & Positive \\
$[I48]$ & 182.0 &  &  & & \\
\hline
ESO - VLT &  289$^\circ$ 35' 49.9' & 400 cm & 0.00664 &  &  \\
(Cerro Paranal, Chile) & -24$^\circ$ 37' 31.5'' & HAWK-I & 0.00667 & V. D. Ivanov & Negative  \\
$[309]$ & 2635.0 &  &  &  & \\
\hline
ESO - La Silla &  289$^\circ$ 15' 58.5' & 355 cm & 0.1 & B. Sicardy &  \\
(La Silla, Chile) & -29$^\circ$ 15' 32.3'' & NTT/SOFI & 0.10003 & D. B\'erard & Negative  \\
$[809]$ & 2345.4.0 &  &  &  & \\
\hline
TRAPPIST-S &  289$^\circ$ 15' 38.2' & 60 cm & 3 & C. Opitom &  \\
(La Silla, Chile) & -29$^\circ$ 15' 16.6'' & TRAPPIST-S & 4.5 & E. Jehin & Negative  \\
$[I40]$ & 2317.7 &  &  &  & \\
\hline
PROMPT & 289$^\circ$ 11' 05.5'' & 40 cm & 4 &  &  \\
(Cerro Tololo, Chile) & -30$^\circ$ 09' 56.3''  & PROMPT & 6 & J. Pollock & Negative \\
$[807]$ & 2225.0 &  & & & \\
\hline
Las Cumbres* &  289$^\circ$ 11' 42.7'' &  100 cm & 3/4 &  &  \\
(Las Cumbres, Chile) & -30$^\circ$ 10' 02.6'' & FLI & 5/6 & F. Bianco & Negative \\
$[W85]$ & 2201.7 & MicroLine 4720 & & & \\
\hline
 &  & Dall Kirkham 50 cm & 3 & A. Maury & Negative \\
San Pedro de Atacama & 291$^\circ$ 49' 13.2'' & Apogee U42 & 4 &  & \\
(S. P. de Atacama, Chile) & -22$^\circ$ 57' 12''.1 &  & & & \\
$[I16]$ & 2398.50 & Ash2 Newtonian 40 cm & 15 & N. Morales & Negative \\
 & & SBIG-STL11000 & 17.59 & & \\
\hline
OPD/LNA & 314$^\circ$ 25' 03.0'' & Perkin Elmer 160 cm & 2 & G. Benedetti-Rossi & \\
(Itajub\'a, Brazil) & -22$^\circ$ 32' 04'' & Andor IXON &  2.950  & B. Morgado & Negative \\
$[874]$ & 1864.0 & DU-888E-C00-\#BV & & A. R. Gomes-Junior & (w/ clouds)
\enddata
\tablecomments{
``FLI'' stands for Finger Lakes Instruments; ``PI'' for Princeton Instruments; ``PL'' for Pro Line. * Three 1-m telescopes were used at Las Cumbres (IAU codes W85, W86 and W87). One of them used I filter to minimize the moon contamination while the other two observed in Clear. Exposure time were 3 seconds for 2 telescopes (filters I and Clear) and 4 seconds for the third telescope. Readout time is not constant for this observation, but in average is 2 seconds. Computers are synchronized via NTP to the site GPS and the start of exposure in the three telescopes had a time shift in order to combine the three data sets and obtain one unique light curve without readout time.}
\end{deluxetable*}

\begin{deluxetable*}{ccccc}
\tablecaption{Observation details of the multi-chord stellar occultation on November 07, 2014 with no data acquired \label{tab:occultation_circumstances_multichord_negative}}
\tablewidth{700pt}
\tabletypesize{\scriptsize}
\tablehead{
\colhead{Site Name} & \colhead{Longitude (E)} & \colhead{Telescope Aperture} &  &  \\
\colhead{(Country)} & \colhead{Latitude (N)} & \colhead{Detector/Instrument} &  \colhead{Observers} & \colhead{Situation} \\
 & \colhead{Altitude (m)} &  &  & 
}
\startdata
Bras\'ilia & 312$^\circ$ 07' 04.02'' & G11 35 cm  &  &  \\
(Brazil) & -15$^\circ$ 47'' 40.37'' & SBIG & P. Cacella & Overcast \\
 & 1096 & ST8XMEI & &  \\
\hline
Vit\'oria (UFES) & 319$^\circ$ 41' 33.18'' & GSO 35 cm & & \\
(Brazil) & -20$^\circ$ 16' 41.43''  & SBIG & M. Malacarne & Overcast \\
 & 6 & ST8XME & \\
\hline
Oliveira (SONEAR) & 315$^\circ$ 10' 27.88''& 45 cm & & \\
(Brazil) &  -20$^\circ$ 42 54.27 & FLI & C. Jacques & Overcast \\
 & 984 & 16803 &  & \\
\hline
U. Aut\'onoma Tom\'as Fr\'ias &  295$^\circ$ 22' 33.00''  & Cassegrain 60 cm &  &    \\
(Bolivia) & -21$^\circ$ 35' 45.90'' & FLI & R. Condori & Overcast \\
 & 1866 & IMG1001E &  &  \\
\hline
S\~ao Carlos (UFSCar) & 312$^\circ$ 06' 50.11'' & Meade LX200 30 cm & & \\
(Brazil) & -21$^\circ$ 58' 49.01'' & SBIG & G. Rojas & Overcast \\
 & 847 & ST9 &  &  \\
\hline
Guaratinguet\'a (UNESP) & 314$^\circ$ 48' 29.50'' & Meade LX200 40 cm & & \\
(Brazil) & -22$^\circ$ 48' 10.02'' & SBIG & R. Sfair & Overcast \\
 & 573 & ST-7XME & & \\
\hline
Rio de Janeiro (ON) & 316$^\circ$ 46' 32,83'' & Meade LX200 30 cm & F. Braga-Ribas &  \\
(Brazil) & -22$^\circ$ 53' 46,85'' & Raptor Merlin 246 & J. I. B. Camargo & Overcast \\
 & 29 &  & & \\
\hline
S\~ao Jos\'e dos Campos (INPE) & 314$^\circ$ 08' 16'' & C11 28 cm &  & \\
(Brazil) & -23$^\circ$ 12' 33'' & SBIG & A. C. Milone  & Overcast \\
 & 975 & ST-7XE & & \\
\hline
Ponta Grossa (UEPG) & 309$^\circ$ 53' 41.39'' & Meade RC40 40 cm & & \\
(Brazil) & -25$^\circ$ 05' 40.27'' & SBIG & M. Emilio & Overcast \\
 & 910 & ST-6303E & \\
\hline
Foz do Igua\c{c}u & 305$^\circ$ 24' 22.55'' & C11 28 cm & & \\
(Brazil) & -25$^\circ$ 26' 05.45'' & SBIG & D. I. Machado & Overcast \\
 & 185 & ST-7XME &  & \\
\hline
CASLEO & 290$^\circ$ 41' 15.9'' & 215 cm & & \\
(Argentina) & -31$^\circ$ 47' 54.7'' & PI-2048B & R. Gil-Hutton & Overcast \\
 & 2552 & & & \\
\hline
Montevideo & 303$^\circ$ 48' 37.1 & Meade LX200 30 cm & G. Tancredi &  \\
(Uruguay) & -34$^\circ$ 45' 20.0'' & Raptor Merlin 246 & S. Roland & Star not detected \\
 & 130.0 &  & L. Almenares & \\
 & & & A. Ceretta &
\enddata
\tablecomments{
``FLI'' stands for Finger Lakes Instruments; ``PI'' for Princeton Instruments.}
\end{deluxetable*}


\subsection{Rotational light curve}
\label{subsec:photometric_observation}

In order to determine the rotational phase of VS2 at the time of the multi-chord occultation, we  performed a photometric follow up of the object a few days after the event. We obtained a total of 97 images of VS2 over three nights -- November 15, 16, and 17, 2014 -- with the 2k $\times$ 2k CCD of the 1.5~m telescope at Sierra Nevada Observatory in Granada (Spain). The image scale of the instrument is 0.232 arcsec/pixel with a Field of View (FOV) of 7.92' $\times$ 7.92'. All the images were acquired without filter and in 2 $\times$ 2 binning mode to maximize SNR. The integration time was 300~s throughout the three nights with a Moon illumination of 36\% during the first night and 18\% during the third night. The average seeing during the three nights was 3.3 arcsec. We pointed the telescope at the same coordinates on the three nights in order to use the same reference stars and thus minimize systematic photometric errors. 

Standard Bias and Flat field corrections were applied on science images. Specific routines written in Interactive Data Language (\textsc{IDL}) were developed to perform the aperture photometry of all the chosen reference stars and VS2. We tried different apertures for the target, calibration stars and sky ring annulus in order to maximize the SNR on the object for each night and to minimize the dispersion of the photometry.

\begin{figure}
\includegraphics[width=\columnwidth]{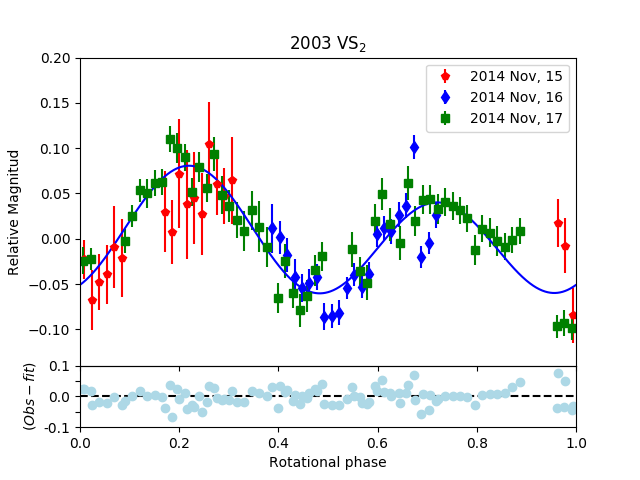}
\caption{Rotational light curve for VS2 obtained on November 15 (in red), 16 (in blue) and 17 (in green), 2014 at Sierra Nevada Observatory (see text for details) just a few days after the occultation. Light blue dots on the bottom panel show the residuals between the observed points and the fit (blue line). The estimated rotational phase for the November 2014 stellar occultation is near one of the brightness maxima. We used the occultation date as the zero phase date.
\label{fig:rotational_light_curve}}
\end{figure}

The flux of VS2 relative to the comparison stars is finally obtained versus the Julian Date, accounting for light travel times. The procedures we used were identical to those detailed in \cite{FernandezV2016,FernandezV2017}. We folded the final photometric data with the well-known rotational period for VS2 obtained in \cite{Santos-Sanz2017}, P= 7.4175285 $\pm$ 0.00001 h. From this rotational light curve, we deduce that VS2 was near one of its absolute brightness maxima at the time of the November 07, 2014 stellar occultation (Fig. \ref{fig:rotational_light_curve}), 
which implies that the object occulted the star when its apparent surface area was near its maximum. These folded data are also fitted with a second order Fourier function in order to obtain the peak-to-valley amplitude of the rotational light curve, which turns out to be $0.141 \pm 0.009$ mag (as listed in Table~\ref{tab:solutions_to_VS2}). The decrease in the peak-to-peak amplitude, compared to the value obtained by \cite{Thirouin2013}, is probably due to geometric effects (i.e. a change in the aspect angle with respect to the previous amplitude estimations). The epoch for zero phase was chosen to be near the time of occultation (November 07, 2014 at 04:00:00 UTC).

\section{Stellar Occultation Data Analysis}\label{sec:analysis}

Some observations involved in the multi-chord event recorded only the integer part of the second in each image header. It was then necessary to retrieve the fractional part of the second for the mid-exposure time of each image by performing a linear fit to the set, see details in \cite{Sicardy2011} and \cite{Dias-Oliveira2017}. From the linear fit we then retrieve the fraction of second for each image with an internal accuracy that depends on the square root of the number of images used, which is less than 0.1~s in practice.

Data obtained in video format from West Park Observatory was converted to Flexible Image Transport System (fits) format using the same procedure as described in \cite{Benedetti-Rossi2016} and \cite{BuieKeller2016}.

The two single-chord events were analyzed to obtain the stellar flux (plus the faint, background contribution due to the occulting body) ratioed to nearby calibration stars using procedures based on the standard Daophot \textsc{IDL} routines \citep{Stetson1987}. The multi-chord occultation was harder to analyze. Since the SNR of the comparison and target stars were too low in all data sets, using the \textsc{IDL} routines the derived chords did not produce satisfactory results, as they presented sizes and positions that did not fit a reasonable and realistic ellipse in the sky plane. We then subjected the data sets to a careful treatment by the new photometry task version of \textsc{PRAIA} -- Program for Reduction of Astronomical Images Automatically \citep{Assafin2011}. 

\begin{figure}
\includegraphics[width=\columnwidth]{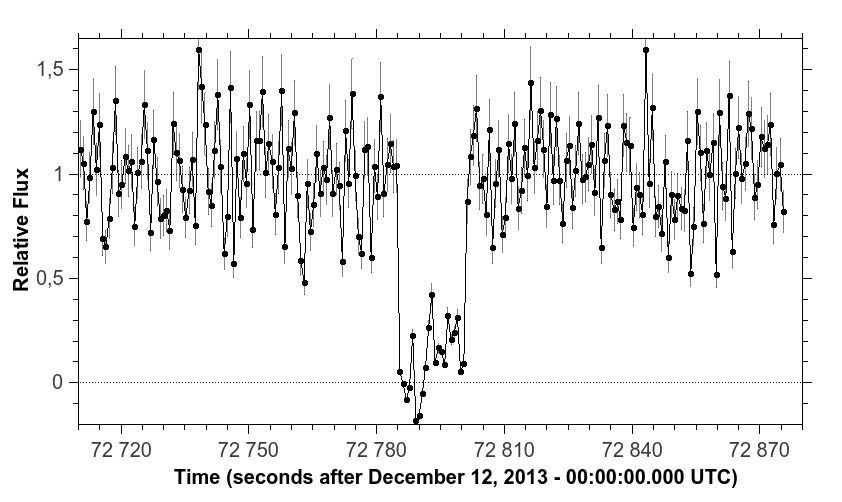}
\includegraphics[width=\columnwidth]{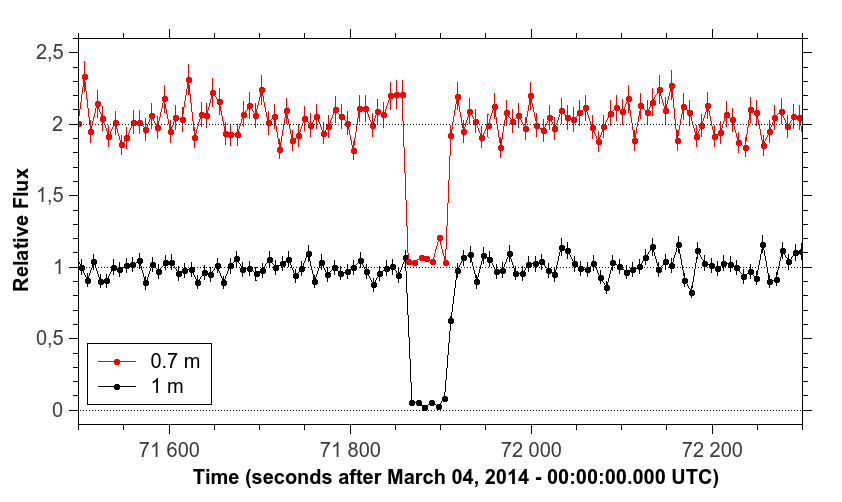}
\includegraphics[width=\columnwidth]{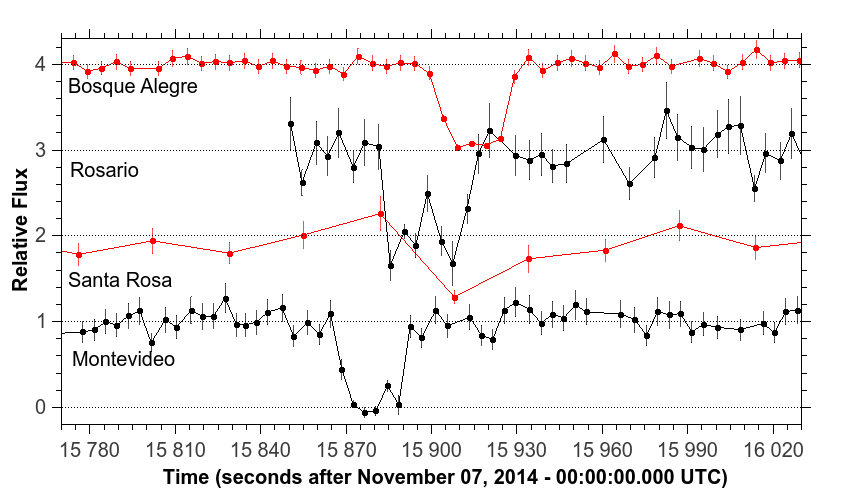}
\caption{Normalized light curves of the positive detections of the stellar occultations -- see Tables~\ref{tab:occultation_circumstances} and \ref{tab:occultation_circumstances_multichord} for sites details. \textbf{Top}: single-chord event in December 12, 2013; \textbf{Center}: single-chord event in March 04, 2014, showing the detection with the two telescopes at the same site; \textbf{Bottom}: multi-chord event on November 2014. Chords are shifted in flux for better visualization. Vertical lines are the uncertainties in photometry for each point.
\label{fig:light_curve_chords}}
\end{figure}

\textsc{PRAIA} uses differential aperture photometry and it differs from the \textsc{IDL} routines in many instances, presenting many useful features which description is beyond the scope of this work. Among them, the apertures, sky ring sizes and widths are automatically optimized for better SNR (outside of the event) for each image, following a thorough aperture centering procedure, and the flux is precisely corrected for the variable apertures for all objects on each image. Calibration fluxes are smoothed and the best set of calibration stars is set automatically as a function of the best light curve standard deviation. The flux ratio vs. time is further corrected by atmospheric systematic factors (using the observed flux before and after the event) by applying a polynomial fit (we used the third degree in all cases). Finally, the flux ratio is normalized and the resulting normalized light curve stored with many kinds of information (errors, photometry parameters, etc). Note that the three stars magnitude are about 16, while it is almost 20 for the TNO. This means that VS2 is not visible in any of the equipment and exposures setup used in the observations and the stars disappear completely during the three occultations, with no residual light from VS2, i.e., we assume that the flux drops to zero in each of these observations. The resulting light curves with positive occultation detections are presented in Fig.~\ref{fig:light_curve_chords}. Light curves flux standard deviation are 21.8\% for the 2013 occultation, 12.9\% and 6.6\% for the two telescopes of 0.7m and 1m, respectively, from the March 2014 event, while it is 6.9\% for Bosque Alegre, 14.7\% for Ros\'ario, 14.5\% for Santa Rosa, and 13.1\% for Montevideo in the multi chord occultation.

The start (star disappearance) and end (star reappearance) times of the occultation are determined by fitting each event by a sharp opaque edge model, after convolution by 
(i) Fresnel diffraction, 
(ii) finite CCD bandwidth, 
(iii) finite stellar diameter, and 
(iv) finite integration time 
(see details in \cite{Widemann2009, Braga-Ribas2013, Ortiz2017}). Note that the Fresnel monochromatic diffraction operates over the Fresnel scale $F= \sqrt{\lambda D/2}$, where $\lambda$ is the wavelength of observation and $D$ the geocentric distance of the object.

The stellar diameter projected at VS2's distance is estimated using the B, V and K apparent magnitudes provided by the NOMAD catalog \citep{Zacharias2004}  and the formulae of \cite{VanBelle1999} -- see Table~\ref{tab:occ_stars_RA_DEC_Gmag}. Using the apparent TNO motion relative to the star, it is possible to express the integration times in actual distances traveled between adjacent data points in the sky plane. Each effect is evaluated in Table~\ref{tab:convolutions}. We note that our data are dominated by the integration times, and not by Fresnel diffraction or stellar diameter. The occultation timings are then obtained by minimizing a classical $\chi^2$ function, using the same procedures as described in \cite{Sicardy2011}.

\begin{deluxetable*}{ccccc}
\tablecaption{Highest amplitude of effects to consider in order to determine the start and end instants of the three stellar occultations \label{tab:convolutions}}
\tablewidth{700pt}
\tabletypesize{\scriptsize}
\tablehead{
\colhead{Occultation Date} & \colhead{D $^1$} & \colhead{Fresnel $^2$} & \colhead{Star Diameter $^3$} & \colhead{Integration Time $^4$}  \\
 & [au] & [km] & [km] & [km]}
\startdata
December 12, 2013 & 35.58 & 1.73 & 0.41 & 18.7 \\
   March 04, 2014 & 36.56 & 1.78 & 0.24 & 36.0 \\
November 07, 2014 & 35.72 & 1.74 & 0.56 & 63.5 \\
\enddata
\tablecomments{
$^1$ - Geocentric distance in astronomical units from Horizons/JPL (\url{https://ssd.jpl.nasa.gov/horizons.cgi}) using JPL\#30 and DE431; 
$^2$ - Fresnel diffraction effect ($F=\sqrt{\lambda D/2}$),
see text;
$^3$ - Stellar diameter projected at VS2's geocentric distance calculated using \citep{VanBelle1999} and assuming super giant stars;
$^4$ - The finite integration time is given for the smallest exposure time of the positive detections. It is expressed in kilometers after accounting for the velocity of the body projected in the sky plane, see text.}
\end{deluxetable*}

\subsection{Single-chord events}
\label{subsec:single_chords_events}

\begin{figure}
\includegraphics[width=\columnwidth]{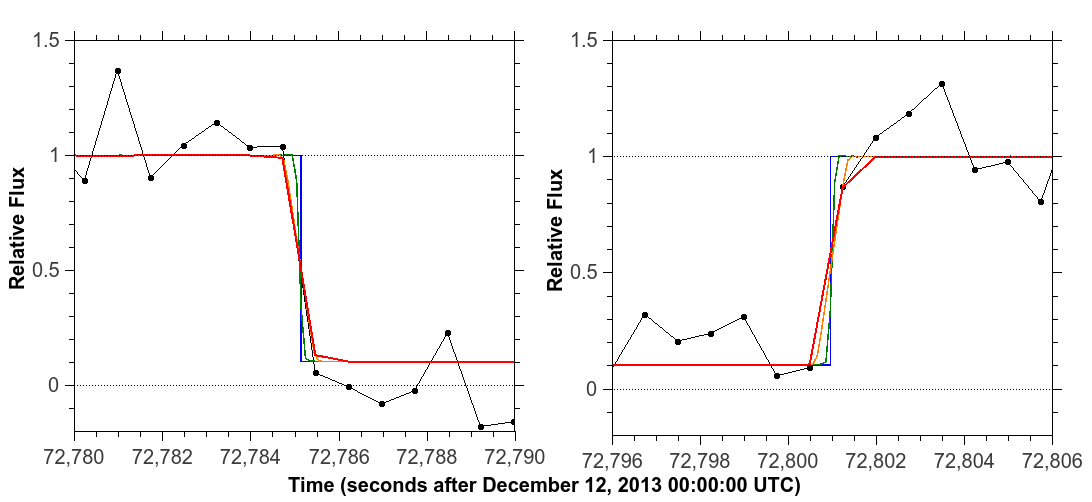}
\includegraphics[width=\columnwidth]{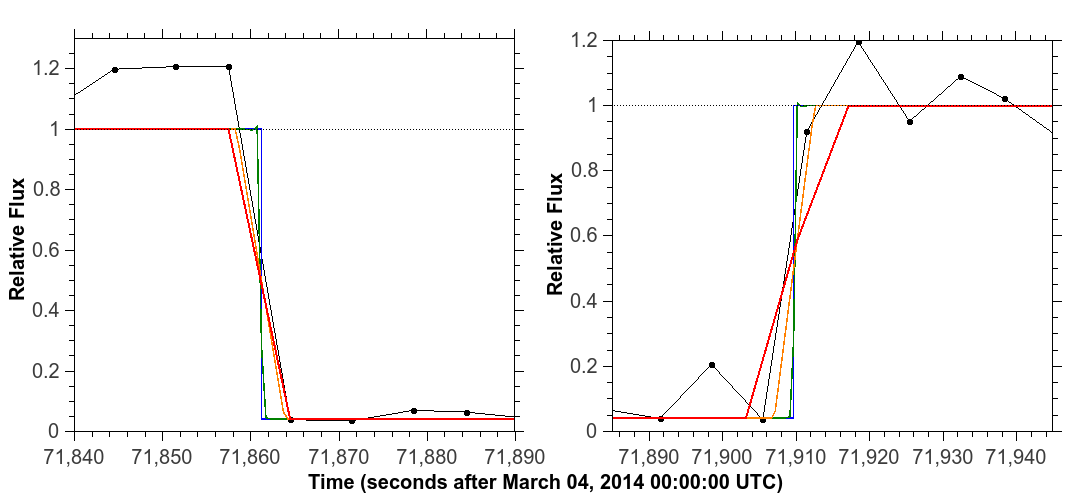}
\includegraphics[width=\columnwidth]{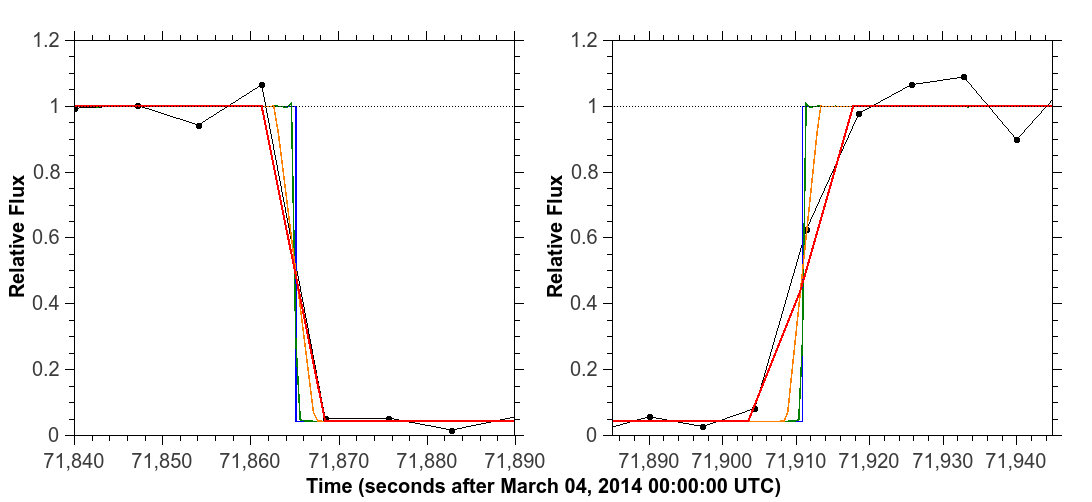}
\caption{Best fit for the star disappearance (left) and reappearance (right) instants of the single-chord events -- see Table~\ref{tab:occultation_circumstances} for sites details. \textbf{Top}: December 12, 2013 occultation; \textbf{Center}: March 04, 2014 occultation observed with the 0.7m telescope; \textbf{Bottom}: March 04, 2014 occultation observed with the 1m telescope. Data points are shown in black, the geometric square well models are in blue, the effect of diffraction is shown in green, the effect of finite integration time is added in the orange, and the red lines show all the effects accounted for, and then used for comparison
with the actual data points.
\label{fig:best_fit_single_chords}}
\end{figure}

The single-chord occultation of December 12, 2013 was detected from La R\'eunion Island and had a shadow velocity of 24.91 km s$^{-1}$, 
while the March 04, 2014 event had a shadow velocity of 8.99 km s$^{-1}$ and it was detected by two telescopes at the same site in Israel (see Table~\ref{tab:occultation_circumstances} for details). The best fits to the ingress and egress profiles are presented in Fig.~\ref{fig:best_fit_single_chords}. The occultation durations of 15.80$^{+0.35}_{-0.40}$~s (La R\'eunion) and 45.75 $\pm$ 1.85~s (Israel) then correspond to occultation chords with lengths of 393.6$^{+8.7}_{-10.0}$ km and 411.3 $\pm$ 16.6 km, respectively, see Table~\ref{tab:single_chord_details}.

The timings in Table~\ref{tab:single_chord_details} provide the extremities of the corresponding occultation chords projected in the sky plane. More precisely, they give the position $(f,g)$ of the star projected in the sky plane and relative to the center of the object. The quantities $f$ and $g$ are expressed in kilometers, and counted positively toward local celestial east and celestial north, respectively. Note that the calculated position $(f,g)$ depends on both the ephemeris used for the body and the adopted star position, that are both determined to within a finite accuracy. As such, $(f,g)$ must be corrected by an offset that can be determined if several chords are available, thus pinning down the position of the object's center position relative to the star.

This is not possible for a single-chord event. In that case, a circle was adjusted to each chord, using the area-equivalent diameter of 564.8 km in order to determine the center of the TNO in the plane of the sky relative to the star at a prescribed time.
Two solutions are then equally possible, depending on whether the center of the body went south or north of the star, as seen from Earth. We consider here the average of the two solutions, still resulting in an accurate astrometric positions for VS2 at the time of the two occultations, with uncertainties smaller than 25 mas for the two single-chord occultations. Note that the diameter used here was obtained from the November 07, 2014 occultation, when the object was at one of the maxima in brightness of the light curve, but we do not know the rotational phase for the other two single-chord occultations. This means that probably the shadow path must be smaller than 564.8 km. Anyway, a difference of 50~km in radius will give an uncertainty of less than 1.7 mas in the object position, which is within the error bars. The object positions in right ascension and declination corrected by the offsets obtained from the occultations are given in Table~\ref{tab:ra_dec_TNO}.

\begin{deluxetable*}{cccc}
\tablecaption{Timings of star disappearances and reappearances \label{tab:single_chord_details}}
\tablewidth{700pt}
\tabletypesize{\scriptsize}
\tablehead{
\colhead{Occultation Event $^{1}$} & \colhead{Disappearance $^{2}$} & \colhead{Reappearance $^{2}$} & \colhead{Chord Size} \\
 & (s) & (s) & (km)}
\startdata
December 12, 2013 & 72785.15$^{+0.15}_{-0.20}$ & 72800.95$^{+0.15}_{-0.25}$ & 393.6$^{+8.7}_{-10.0}$ \\
\\
March 04, 2014 (1m) & 71865.15 $\pm$ 1.55 & 71910.9$\pm 0.3$ & 411.3 $\pm$ 16.6 \\
\\
March 04, 2014 (70cm) & 71861.25 $\pm$ 1.15 & 71909.7$^{+0.5}_{-0.6}$ & 435.6$^{+14.8}_{-15.7}$\\
\\
November 07, 2014 (BA) & 15903.354 $\pm$ 0.646 & 15927.400 $\pm$ 0.642 & 517.7 $\pm$ 27.8 \\
\\
November 07, 2014 (RO) & 15883.351 $\pm$ 0.887 & 15913.006 $\pm$ 0.670 & 638.5 $\pm$ 33.5 \\
\\
November 07, 2014 (SR) & 15901.0 $\pm$ 8.0 & 15928.0 $\pm$ 8.0 & 581 $\pm$ 345 \\
\\
November 07, 2014 (MO) & 15868.286 $\pm$ 0.800 & 15890.583 $\pm$ 0.800 & 480.1 $\pm$ 34.4 \\
\enddata
\tablecomments{
BA: Bosque Alegre; RO: Ros\'ario; SR: Santa Rosa; MO: Montevideo;
$^1$ - See Tables \ref{tab:occultation_circumstances}, \ref{tab:occultation_circumstances_multichord}, and \ref{tab:occultation_circumstances_multichord_negative} for occultation details.
$^2$ - Times are given in seconds after 00:00:00.000 UTC of the occultation day.}
\end{deluxetable*}

\begin{deluxetable*}{ccccc}
\tablecaption{Astrometric right ascention and declination, and the offsets ($f,g$) for VS2 derived from the three stellar occultations
\label{tab:ra_dec_TNO}}
\tablewidth{700pt}
\tabletypesize{\scriptsize}
\tablehead{
\colhead{Occultation Date} & \colhead{RA ICRS} & \colhead{DEC ICRS} & \colhead{f *} & \colhead{g *} \\
 & \multicolumn{2}{c}{(JPL30 and DE431 + Offset)} & (mas) & (mas)}
\startdata
December 12, 2013 20:13:00.000 & 04$^h$ 36$^{m}$ 19$^{s}$.83798 & +33$^\circ$ 55' 55''.2525 & -15.7 $\pm$ 2.4 & -17.5 $\pm$ 8.1 \\
\\
March 04, 2014 19:55:00.000 & 04$^h$ 32$^{m}$ 04$^{s}$.47805 & +33$^\circ$ 24' 42''.0523 & -12.2 $\pm$ 5.4 & -23.7 $\pm$ 6.1 \\
\\
November 07, 2014 04:22:00.000 & 04$^h$ 48$^{m}$ 32$^{s}$.13788 & +33$^\circ$ 58' 36''.1927 & -60.1 $\pm$ 0.3 & -24.5 $\pm$ 0.4 \\
\enddata
\tablecomments{
* The offsets depend also on the adopted star positions given on Table~\ref{tab:occ_stars_RA_DEC_Gmag}.}
\end{deluxetable*}

\subsection{Multi-chord event\label{subsec:multi_chords_event}}

The November 07, 2014 occultation had a typical shadow velocity of 21.53 km s$^{-1}$ and observers attempted to record the event from 21 different sites. Unfortunately, due to bad weather only eight sites were successful in gathering data, among which four detected the event (see Tables \ref{tab:occultation_circumstances_multichord} and \ref{tab:occultation_circumstances_multichord_negative}). Despite non-detections, the chords obtained at Cerro Tololo (PROMPT), Las Cumbres, and La Silla (NTT/SOFI; Son of ISAAC
\citep{Moorwood1998, Moorwood1998b}) provide important constraints on the size of the object, 
being close to the occultation path.

\begin{figure}
\includegraphics[width=\columnwidth]{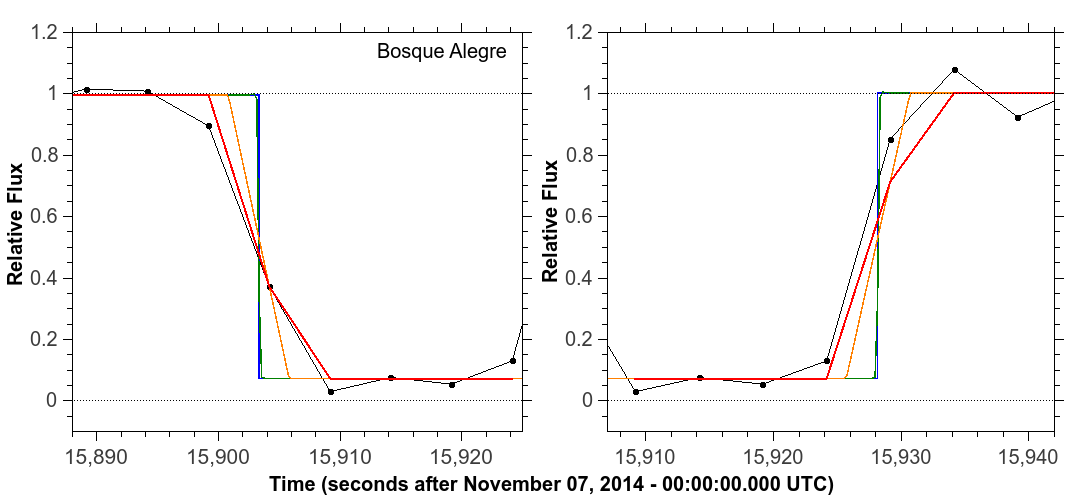}
\includegraphics[width=\columnwidth]{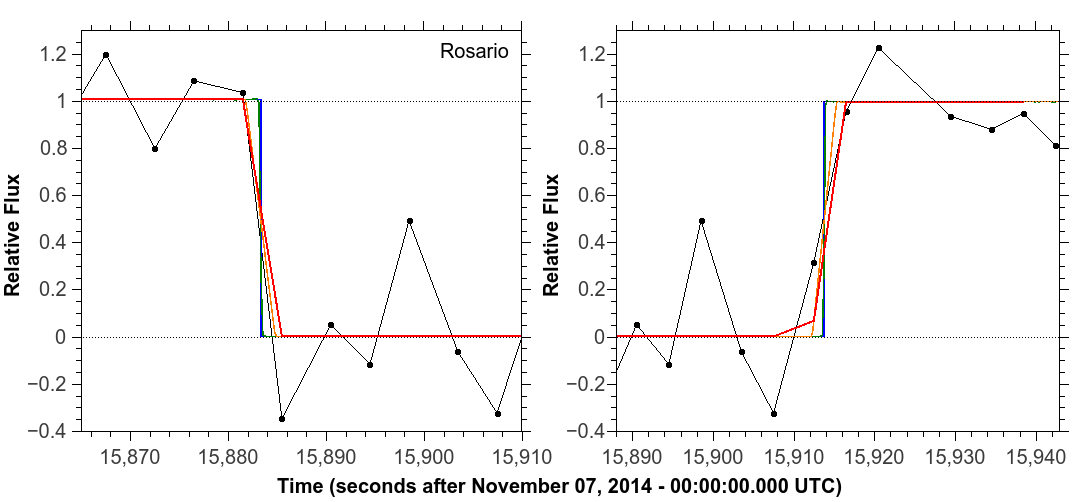}
\includegraphics[width=\columnwidth]{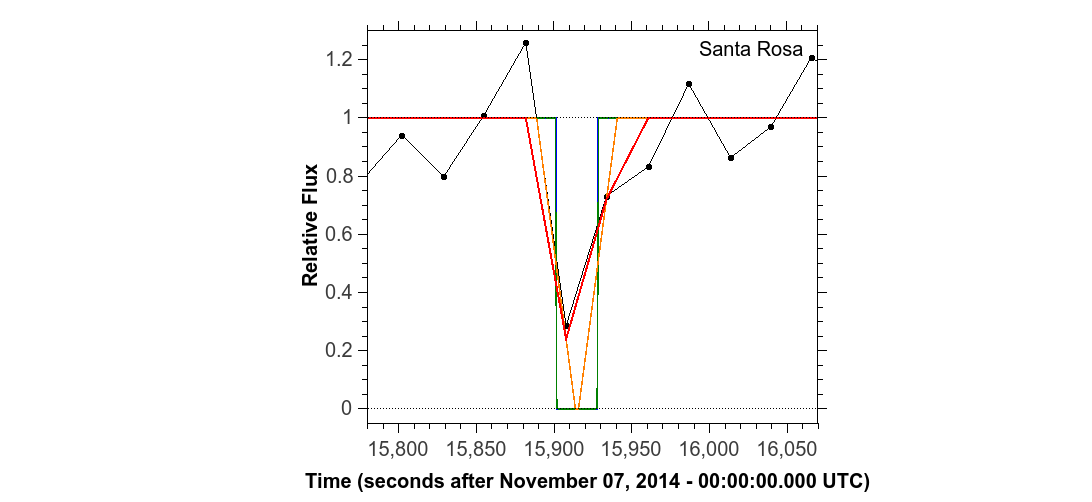}
\includegraphics[width=\columnwidth]{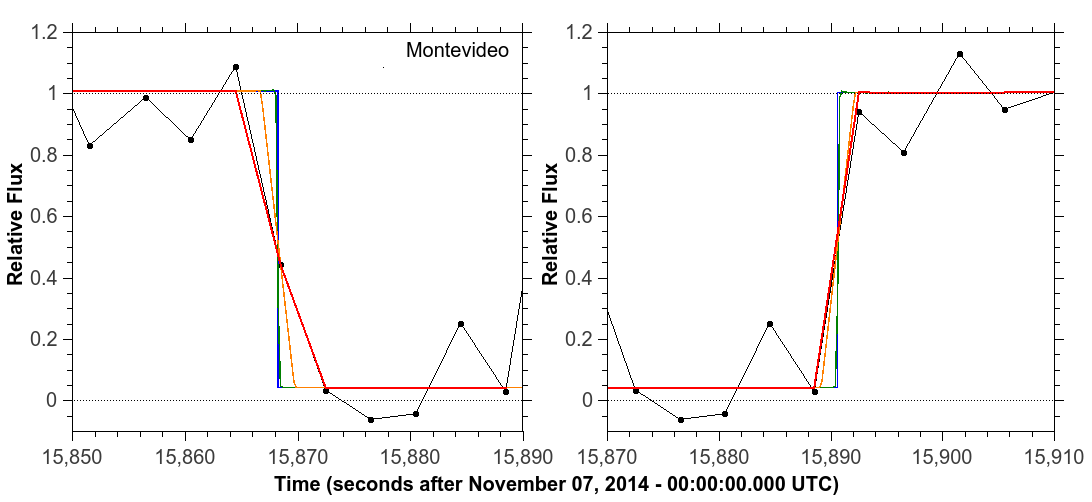}
\caption{Best fit for the star disappearance (left) and reappearance (right) instants of the multi-chord event of November 07, 2014 -- see Table~\ref{tab:occultation_circumstances_multichord} for sites details. From top to bottom: Bosque Alegre, Rosario, Santa Rosa, and Montevideo. Line colors are the same as in Fig. \ref{fig:best_fit_single_chords}.
\label{fig:best_fit_multi_chords}}
\end{figure}

The resulting light curves with positive detection, shown in Fig.~\ref{fig:light_curve_chords}, provide $N=8$ chord extremities, from which we derived the disappearance and reappearance instants of the occultation (see Table~\ref{tab:single_chord_details}). 
The best fits for the ingress and egress times are presented in Fig.~\ref{fig:best_fit_multi_chords}.


The  shape of the body's limb is assumed to be an ellipse, as adopted in previous works, see 
\cite{Braga-Ribas2013, Braga-Ribas2014, Benedetti-Rossi2016, Dias-Oliveira2017,Ortiz2017}. The ellipse is defined by $M=5$ adjustable parameters: 
the position of the ellipse center, $(f_{\rm c},g_{\rm c})$; 
the apparent semi-major axis, $a'$; 
the apparent oblateness, $\epsilon' = (a'-b')/a'$ 
(where $b'$ is the apparent semi-minor axis); 
and the position angle of the pole $P'$ of $b'$, 
which is the apparent position angle of the pole measured eastward from celestial north. 
Note that the center $(f_{\rm c},g_{\rm c})$ actually measures the offsets in right ascension and declination to be applied to the adopted ephemeris, assuming that the star position is correct. Note also that the quantities $a'$, $b'$, $f_{\rm c}$, and $g_{\rm c}$ are all expressed in kilometers.

The statistical significance of the fit is evaluated from the $\chi^2$ per degree of freedom (pdf) defined as 
$\chi^{2}_{pdf}$ = $\chi^2/(N - M)$, 
which should be close to unity for a correct fit. 
The individual 1-$\sigma$ error bar of each parameter is obtained by varying that parameter from its nominal solution value (keeping the other parameters free), 
so that $\chi^2$ varies from its minimum value 
$\chi^{2}_{\rm min}$ to $\chi^{2}_{\rm min}$ + 1.

\begin{figure}
\includegraphics[width=\columnwidth]{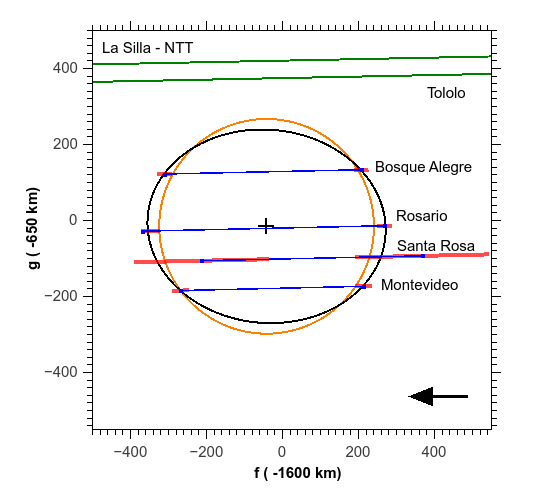}
\caption{Occultation chords (blue) and their uncertainties (red) from the multi-chord event of November 2014 with the best elliptical fit (black). The offsets in both axis are with respect to ephemeride positions obtained using JPL\#30 and DE431. The arrow shows the direction of the shadow motion and the green lines represent the two closest sites from where there were no detection of the occultation 
(see Table~\ref{tab:occultation_circumstances_multichord}). The equivalent circle (that has the same area as the ellipse) is shown in orange with a diameter of 564.8 km.
\label{fig:fit_ellipse}}
\end{figure}

Using the timings for the star disappearance and reappearance obtained from the four positive detections
(Table~\ref{tab:single_chord_details}),
we obtain a best-fitting ellipse with $\chi^2_{pdf}=0.78$ and its parameters
$a' = 313.8 \pm 7.1$ km, 
$\epsilon' = 0.190^{+0.052}_{-0.060}$
($b' = 254.8^{+25.0}_{-21.7}$),
$(f,g) = (1558.1 \pm 8.1, 634.6 \pm 11.0)$ km and 
$P' =  5 \pm 7^{\circ}$, 
with equivalent radius 
$R_{\rm eq} = 282.4^{+16.9}_{-15.1}$ km, 
as shown in Fig.~\ref{fig:fit_ellipse}.

\section{Results}\label{sec:results}

\subsection{Size and Shape}
\label{subsec:shape}

Homogeneous rocky objects with diameters on the order of 
1000~km or more are expected to have reached the hydrostatic equilibrium, assuming Maclaurin spheroidal or Jacobi ellipsoidal shapes \citep{Chandrasekhar1987}.
The critical diameter allowing to reach this equilibrium is discussed by \cite{TancrediFavre2008}. 
In particular, hydrostatic equilibrium could be reached for smaller objects, if made of ices or of a combination of ices and rocks. Nevertheless, the formation scenario and collisional history of individual objects are not known, allowing objects with complex internal structures (i.e., differentiated density layers, granular for a peeble accretion scenario, etc.).
Thus, we cannot discard solutions that diverge somewhat from the expected equilibrium figure given by \cite{Chandrasekhar1987}. The case of the dwarf planet Haumea \citep{Ortiz2017} is a good example of a body with sizes that does not match a hydrostatic equilibrium figure for a geologically homogeneous fluid body. 

The rotational light curve of Fig. \ref{fig:rotational_light_curve} shows that VS2 is consistent with an object having a triaxial ellipsoidal shape. The short-term photometry of the body presented in Section~\ref{subsec:photometric_observation} showed that the multi-chord occultation took place near one of the absolute brightness maxima of the rotational light curve, meaning that VS2 occulted the star when its apparent surface area was near its maximum. Assuming a triaxial shape, this occurs when its longest axis $a$ was perpendicular to the line of sight, so that the semi-major axis of the projected ellipse $a'$ is equal to the real 3D-axes. With the known rotation period and phase, we verified if it was possible to use one or the two single-chord occultation results to improve the accuracy on the determination of size and shape for the TNO. The problem is that even if the rotational phase and period are well determined, the ephemeris for the object is not well constrained, with uncertainties of about 50 mas, in a way that the chord from the two occultations can fit anywhere in the multi-chord apparent ellipse. With this uncertainty we could not use any of the other chords to obtain further constraints on VS2's shape using the same method that \cite{Dias-Oliveira2017} did for the TNO 2003~AZ$_{84}$.

Considering the limb fit to the apparent ellipse obtained from the multi-chord occultation and the light curve amplitude, we can use the same procedure as in \cite{Sicardy2011} (also used in \cite{Braga-Ribas2013} and \cite{Ortiz2017}) to derive the three axes of the body. Our best solution for a triaxial shape has semi-axis values
$a = 313.8 \pm 7.1$ km,
$b = 265.5 ^{+8.8}_{-9.8}$ km ($\beta = 0.846 ^{+0.031}_{-0.028}$), 
$c = 247.3 ^{+26.6}_{-43.6}$ km ($\gamma = 0.788 ^{+0.085}_{-0.139}$), with the c-axis inclined $\theta = 65 ^{+15}_{-10}~^\circ$ with respect to the observer, which is not consistent with the Jacobi equilibrium figure -- see Appendix~\ref{AppendixA} for details. This solution gives a spherical volume equivalent diameter of $548.3 ^{+29.5}_{-44.6}$, from which we can derive (using H$_v$ = 4.130 $\pm$ 0.070 mag) the geometric albedo
$p_V = 0.131 ^{+0.024}_{-0.013}$.

Since we can not make any constraint about VS2's density using the triaxial approach without making any other assumption of its mass, we can make an assumption that VS2 has an oblate shape, as a Maclaurin equilibrium figure with semi-axis $a = 313.8$ km and $254.8 \le c \le 313.8$ km. Note that with this assumption, we are also assuming that the light curve amplitude variation is due to albedo features. Making this assumption and combining with the rotation period (given in Table~\ref{tab:previous_publications}), we can derive its density of $1400^{+1000}_{-300}$ kg m$^{-3}$. All those values are summarized in Table~\ref{tab:solutions_to_VS2}.

\begin{deluxetable*}{cc}
\tablecaption{Solution for 2003~VS$_2$ derived from the multi-chord stellar occultation and the rotational light curve. \label{tab:solutions_to_VS2}}
\tablewidth{700pt}
\tabletypesize{\scriptsize}
\tablehead{
\colhead{Parameter} & \colhead{Value}}
\startdata
$a' \times b'$ (km)  & $313.8 \pm 7.1 \times 254.8^{+25.0}_{-21.7}$\\
\\
($f,g$) (km) $^1$ & (-1558.1 $\pm$ 8.1 ; -634.6 $\pm$ 11.0)  \\
\\
Position angle of projected ellipse (deg) & $5 \pm 7$  \\
\\
Light curve amplitude $^2$ -- $\Delta m$ -- (mag) & 0.141 $\pm$ 0.009 \\
\\
Geometric albedo -- $p_V$ & $0.131 ^{+0.024}_{-0.013}$.  \\
\\
$a \times b \times c$ (km) & $313.8 \pm 7.1 \times 265.5 ^{+8.8}_{-9.8} \times 247.3 ^{+26.6}_{-43.6}$  \\
\\
Aspect angle (deg) & $65 ^{+15}_{-10}$ \\
\\
Equivalent diameter (km) & $548.3^{+29.5}_{-44.6}$   \\
\\
Density -- Maclaurin (kg m$^{-3}$) &  $1400^{+1000}_{-300}$
\enddata
\tablecomments{
$^1$ Offsets obtained with respect to JPL30 + DE431. It also depends on the star position given in Table~\ref{tab:occ_stars_RA_DEC_Gmag}.$^2$ Period used to fit was 7.4175285 $\pm$ 0.00001 h from \cite{Santos-Sanz2017}.}
\end{deluxetable*}

\subsection{Atmosphere and secondary detections}\label{subsec:atmosphere}

On the largest TNOs the most common volatiles found through spectroscopic studies are water ice (H$_{2}$O), methane (CH$_4$), ammonia (NH$_3$), molecular nitrogen (N$_2$), and even methanol (CH$_{3}$OH) and ethane (C$_2$H$_6$) for a few objects \citep{Guilbert2009,Barucci2011}. In the case of VS2, near infrared spectra of this body shows the presence of exposed water ice \citep{Barkume2008}, but no other volatile is reported to be detected. Data from the three stellar occultations showed no compelling evidence for a global atmosphere.

Using the same method for Quaoar \citep{Braga-Ribas2013}, we have modeled synthetic occultation light curves using a ray tracing code as described in \cite{Sicardy2011} and \cite{Widemann2009}. Several synthetic light curves are then compared to our best light curve obtained at Wise Observatory on the single chord occultation on March 04, 2014 providing a range of $\chi^2$ with the detection threshold. We test two possible atmospheric models. one with an isothermal N$_2$ atmosphere, as this gas is the most volatile one among those listed above. We consider a typical temperature of 40 K for this model, but the result is weakly dependent on this particular value. The other model is a N$_2$ atmosphere with CH$_4$ as a minor species, starting near 40 K at the surface and ramping up to 100 K near 25 km altitude due to methane IR absorption, as is observed in Pluto's atmosphere \citep{Hinson2017}.

Our $\chi^2$-tests show that for both models, the Wise data are consistent with no atmosphere, with upper limits of 0.2 microbar and 1 microbar at the 1$\sigma$ and 3$\sigma$ levels, respectively.


\begin{figure}
\includegraphics[width=\columnwidth]{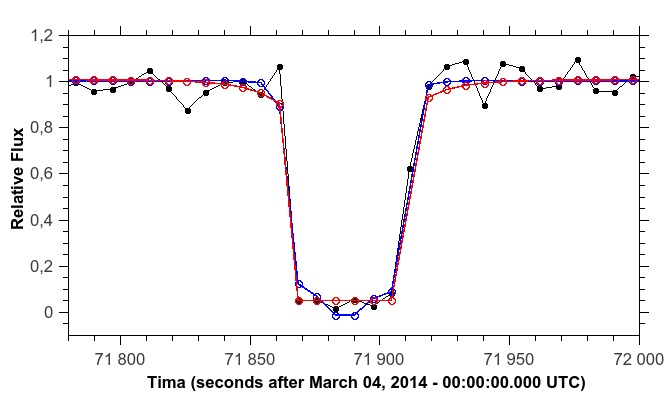}
\caption{Light curve from the single-chord stellar occultation of March 04, 2014 obtained at Wise Observatory (black) and the two models for the isothermic pure nitrogen (red) and a Pluto-like (blue) atmospheres within a 3$\sigma$ limit.
\label{fig:atmosphere}}
\end{figure}

We also searched for secondary events that may be related to a satellite, ring, or some material orbiting VS2. Using our highest time-resolution light curve -- obtained with the 3.5m NTT telescope with 0.1 seconds of cycle time and roughly positioned 400 km from VS2's center -- we can estimate an upper limit of 0.2 km for an opaque object (or 18.5 km for any material with optical depth of 0.1). Considering the best chord with positive detection -- from Bosque Alegre with 5 seconds cycle time -- those limits are 107.7 km for opaque and 732 km for material with optical depth of 0.1.

\begin{figure}
\includegraphics[width=\columnwidth]{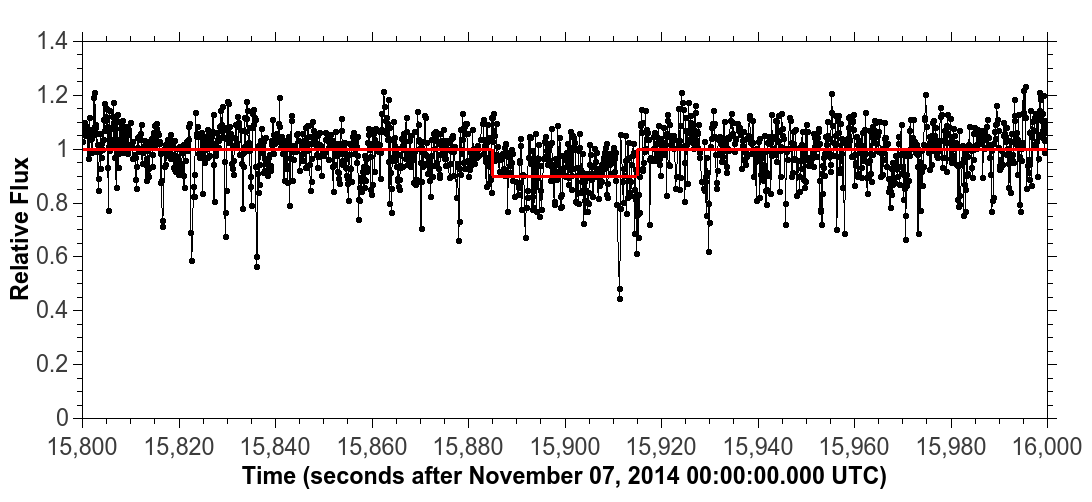}
\caption{Light curve from the multi-chord stellar occultation of November 07, 2014 obtained at La Silla/NTT. All the flux drops in the curve are due to the star being too close to the edge of the CCD or being influenced by the bad row of pixels in the center of the CCD. Note the relative flux drop of 0.09 that happens between instants 15885 s and 15915 s, fitted using a square well model, represented by the red line.
\label{fig:lc_secondary_event}}
\end{figure}

One secondary feature was also detected in the NTT light curve, as shown in Fig.~\ref{fig:lc_secondary_event}. The drop has a relative depth of less than 10\% (compared to the light curve standard deviation of 8.6\%) and a duration of about 29~s, corresponding to a chord length of 625~km. However, this drop remains marginally significant and cannot be confirmed or rejected using the nearby chords from Cerro Tololo (insufficient SNR) and Pico dos Dias (overcast). 
This drop could be due to an occultation of a companion star by the object, causing a secondary occultation as happened during a stellar occultation by the Centaur object Chariklo in 2014 -- see \cite{Leiva2017} and \cite{Berard2017}. Note that with the size of VS2 and the geometry of the occultation, the Bosque Alegre positive light curve (data set with the best SNR) should not present any sign of this possible secondary occultation. Also, considering the main star magnitude (V=15.82) and the drop of 10\%, the companion would have a magnitude V=18.21. If this star is closer than 0.8~arcsec it is likely not to be present in the GDR2 catalog \citep{Brandeker2019}. We estimate that the companion star should be closer than 0.5~arcsec, so the existence of a stellar companion remains consistent with its absence in the GDR2.

Another possibility is that this drop is due to some diffuse material around VS2, but since none of the other light curves show such secondary event to within the 5-$\sigma$ level, this hypothesis is less likely. Also, a careful look on the NTT images presents some issues. The images have 320 vs. 32 pixels with a dead row in the middle of the CCD (row 69). There was only one reference star in the field for photometric calibration, which is fainter than the occulted star, and the telescope tracking was not perfect, in a way that sometimes one or both stars were close to the edge of the CCD. So we can not completely discard instrumental effects or reduction artifacts on this data set. In conclusion, more occultations are needed in order to confirm or discard the presence of any secondary feature.

\section{Conclusions}\label{sec:conclusions}

We observed two single-chord and one multi-chord stellar occultations by the plutino (84922) 2003~VS$_2$. Observations were performed on December 12, 2013 in La R\'eunion, on March 04, 2014 in Israel, and from 11 sites in South America on November 07, 2014, with 4 positive detections. The two single-chord events are consistent with the multi-chord solutions but do not bring further constraints to the object shape. However, they provide good astrometric constraints, thus improving the ephemeris of the body.

Using the timings of the November 07, 2014 event, we find the apparent ellipse with area equivalent diameter of 564.8$^{+33.8}_{-30.2}$ km.
Assuming that VS2 has a triaxial shape and combining the occultation data with the light curve amplitude ($\Delta m = 0.141 \pm 0.009$) obtained from photometric observations of the TNO, we derive the object 3D-shape, with semi-axes 
$a = 313.8 \pm 7.1$ km, 
$b = 265.5^{+8.8}_{-9.8}$ km, 
$c = 247.3^{+26.6}_{-43.6}$ km, with the c-axis inclined
$\theta = 65^{+15}_{-10} ~^\circ$ with respect to the observer. Those values are not consistent with a Jacobi equilibrium figure, but are in accordance, at the 1-$\sigma$ level, with previous values published by \cite{Mommert2012}, \cite{Stansberry2008}, and \cite{Lellouch2013} (area equivalent diameter of 523$^{+35.1}_{-34.4}$ km).

We also considered the possibility that VS2 is an oblate Maclaurin equilibrium figure (with $a=b>c$). With this assumption, the light curve amplitude variation is due to albedo features and VS2 would have a density of $\rho = 1400^{+1000}_{-300} $~kg~m$^{-3}$. The light curve amplitude can be explained in the extreme cases to be due only to the shape (with the triaxial shape model) or to a darker or brighter spot on the surface of about 100 km -- 16\% of object diameter -- if VS2 is a Maclaurin.

A possible secondary event was detected in the NTT data, that may be caused by the presence of some feature in VS2's surroundings or due to a secondary occultation occurred by the presence of a companion star, but the data are insufficient to make any deep conclusions. We also derived an upper limit for a global atmosphere (either isothermic pure nitrogen or Pluto-like) of 1 microbar (3$\sigma$).

Although the rotation period is well defined, and values for the geometric albedo and size of VS2 are well determined with the occultations, its 3D-shape (and as a consequence, its density) still need more observations in order to constrain the values obtained in this work.

\acknowledgments

G.B.-R. is thankful for the support of the CAPES and FAPERJ/PAPDRJ (E26/203.173/2016) grant.  Part of the research leading to these results has received funding from the European Research Council under the European Community’s H2020 (2014-2020/ERC Grant Agreement no. 669416 ``LUCKY STAR''). The research leading to these results has received funding from the European Union’s Horizon 2020 Research and Innovation Programme, under Grant Agreement No. 687378 (SBNAF). P.S.-S. and J.L.O. acknowledge the financial support by the Spanish grant AYA-2017-84637-R and the Proyecto de Excelencia de la Junta de Andaluc\'ia J.A. 2012-FQM1776. P.S.-S., J.L.O. and R.D. acknowledges financial support from the State Agency for Research of the Spanish MCIU through the “Center of Excellence Severo Ochoa” award for the Instituto de Astrof\'isica de Andaluc\'ia (SEV-2017-0709). Based on observations made with ESO Telescopes at the La Silla Paranal Observatory under programme ID 094.C-0352. MA thanks CNPq (Grants 427700/2018-3, 310683/2017-3, and 473002/2013-2) and FAPERJ (Grant E-26/111.488/2013). J.I.B.C. acknowledges CNPq grant 308150/2016-3. R.V.-M. thanks grants: CNPq-304544/2017-5, 401903/2016-8, Faperj: PAPDRJ-45/2013 and E-26/203.026/2015. F.B.-R. acknowledges CNPq grant 309578/2017-5. E.F.-V. acknowledges UFC 2017 Preeminent Postdoctoral Program (P$^3$). TRAPPIST-South is a project funded by the Belgian Fonds (National) de la Recherche Scientifique (F.R.S.-FNRS) under grant FRFC 2.5.594.09.F. E.J is a FNRS Senior Research Associate. A.A.C acknowledges support from FAPERJ (grant E-26/203.186/2016) and CNPq grants (304971/2016-2 and 401669/2016-5). B.M. thanks the CAPES/Cofecub-394/2016-05 grant. A.R.G.J. and R.S thanks the financial support of FAPESP (proc. 2018/11239-8, proc. 2011/08171-3, proc. 2016/ 24561-0). A.M. thanks Caisey Harlingten for the use of his 50cm telescope. We thank V. Buso and R. Condori for his observation efforts.



\appendix

\section{Deriving VS2's shape and Density \label{AppendixA}}
We describe here the procedure used to determine the values of the semi-axis of a triaxial ellipsoid ($a > b > c$), and the angle between the c-axis and the observer ($\theta$).

First we have the relations from \cite{Sicardy2011} (also used in \cite{Braga-Ribas2013} and  \cite{Ortiz2017}) between the real $c$ and $b$ semi-minor axis and the observed $b'$:

\begin{equation}
\label{eq:bprime_original}
    b' = \sqrt{ c^2 ~sin^2 \theta + b^2 ~cos^2 \theta },
\end{equation}

where in our case we have $b' = 254.8 ^{+25.0}_{-21.7}$ km as the semi-minor axis of the ellipse observed in occultation. Multiplying both sides by $1/a$ and replacing $b/a$ by $\beta$, and $c/a$ by $\gamma$, we can rewrite eq.~\ref{eq:bprime_original} and rearrange the terms to have a direct relation of $\gamma$ as a function of $\beta$:

\begin{equation}
\label{eq:gamma_bprime}
    \gamma^2 =
    {\frac{\beta'^2 - \beta^2 cos^2 \theta}{sin^2 \theta}}
\end{equation}

Note that since the multi-chord occultation took place near the maximum brightness, we have $a' = a$, and so $\beta' = b'/a$. Second, using equation (8) from \cite{Sicardy2011} (SI) we have

\begin{equation}
\label{eq:lc_amplitude_vs_abc_axis}
    \Delta m = -1.25 \cdot \log_{10}{ 
    \left[
     \frac{1 + \gamma^2 \tan^{2}{\theta} }{ 1 + ( \gamma / \beta )^2 \tan^{2}{\theta} } 
    \right],
    }
\end{equation}

where $\Delta m$ is the amplitude of the rotational light curve (0.141 $\pm$ 0.009 mag -- see Table~\ref{tab:previous_publications}). We can rearrange the terms in the equation and obtain another relation of $\gamma$ as function of $\beta$:

\begin{equation}
\label{eq:gamma_deltam}
    \gamma^2 = 
    \frac{\beta^2 ~(\xi -1)}{tg^2 \theta ~(\beta^{2} - \xi)},
\end{equation}

where $\xi = 10^{- \Delta m/1.25}$. Those two equations combined (eqs.~\ref{eq:gamma_bprime} and \ref{eq:gamma_deltam}) give us constraints on the values $\beta$ for every $\theta$, but without any further assumptions, $\gamma$ can assume any value from zero (or more specifically, undefined) to $\beta$ and $\theta$ can have values from $0$ to $90$ degrees.

Now, assuming that VS2 is large enough to achieve hydrostatic equilibrium, we have the limits for $\beta$ and $\gamma$ as a Jabobi-shape object. \cite{TancrediFavre2008} present the relations $\Gamma$ -- associated with the angular momentum $L$ -- and $\Omega$ -- associated with the angular velocity $\omega$ -- given respectively by:

\begin{equation}
\label{eq:tancredifavre1}
    \Gamma = \frac{L}{\sqrt{G M^3 R}},
\end{equation}

and

\begin{equation}
\label{eq:tancredifavre2}
    \Omega = \frac{\omega^2}{\pi G \rho },
\end{equation}

where $G$ is the gravitational constant, $M$, $R$, and $\rho$ are the mass, equivalent radius and density of the body, respectively. For a Jacobi object, there are a lower and an upper limit for the two quantities: $0.303 \leq \Gamma \leq 0.390$, and $0.284 \leq \Omega \leq 0.374$; which also limit the values of $\beta$ between 0.432 and 1, and $\gamma$ between 0.345 and 0.583. The shapes of Jacobi ellipsoids in terms of the semi-axes ($a, b$, and $c$) can only be obtained by solving \cite{Chandrasekhar1969} -- also in \cite{LacerdaJewitt2007} and \cite{Sicardy2011} Suplemmentary Information -- integrals:

\begin{equation}
\label{eq:chandrasekhar_integral_beta_gamma}
    \beta^2 \int_0^\infty \frac{du}{(1+u) ~(\beta^2 +u) ~\Delta(\beta,\gamma,u)}
    =
    \gamma^2 \int_0^\infty \frac{du}{(\gamma^2+u) ~\Delta(\beta,\gamma,u)}
\end{equation}

and

\begin{equation}
\label{eq:chandrasekhar_integral_omega}
    \frac{\omega^2 a^3}{G M}
    =
    \frac{3}{2} \int_0^\infty  \frac{u ~du}{(1+u) ~(\beta^2+u) ~\Delta(\beta,\gamma,u)},
\end{equation}

where $\Delta(\beta,\gamma,u) = [(1 + u) ~(\beta^2 + u) ~(\gamma^2 + u)]^{1/2}$. Once $\beta=b/a$ is given, Eq.~\ref{eq:chandrasekhar_integral_beta_gamma} yields $\gamma$, which in turn allows to calculate $\omega$ (or $\rho$), using Eq.~\ref{eq:chandrasekhar_integral_omega}.

\begin{figure}
\begin{center}
\includegraphics[width=.8\columnwidth]{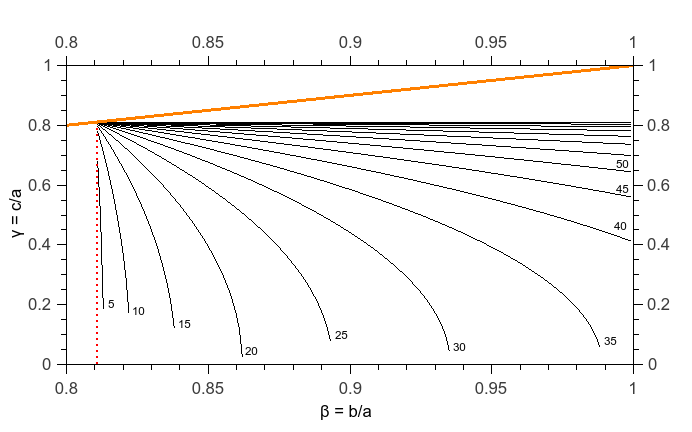}
\caption{Relation between $\beta = b/a$ and $\gamma = c/a$ obtained using eq.~\ref{eq:gamma_bprime}. Black lines correspond to a value of $\theta$ between 5 and 90 degrees (from left to right), every 5 degrees. Orange line are the values for $\beta = \gamma$. Red dotted vertical line is the value of $\beta'=0.811$, observed in occultation.
\label{fig:beta_gamma_lellouch_beta_prime}}
\end{center}
\end{figure}

\begin{figure}
\begin{center}
\includegraphics[width=.8\columnwidth]{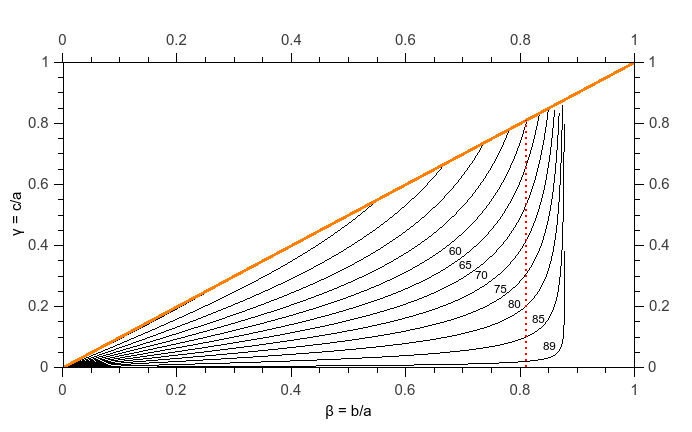}
\caption{Relation between $\beta = b/a$ and $\gamma = c/a$ obtained using eq.~\ref{eq:gamma_deltam}. Black lines correspond to a value of $\theta$ between 0 and 90 degrees (from left to right), every 5 degrees. Orange line are the values for $\beta = \gamma$. Red dotted vertical line is the value of $\beta'=0.811$, observed in occultation.
\label{fig:beta_gamma_sicardy_delta_m}}
\end{center}
\end{figure}

\begin{figure}
\begin{center}
\includegraphics[width=.8\columnwidth]{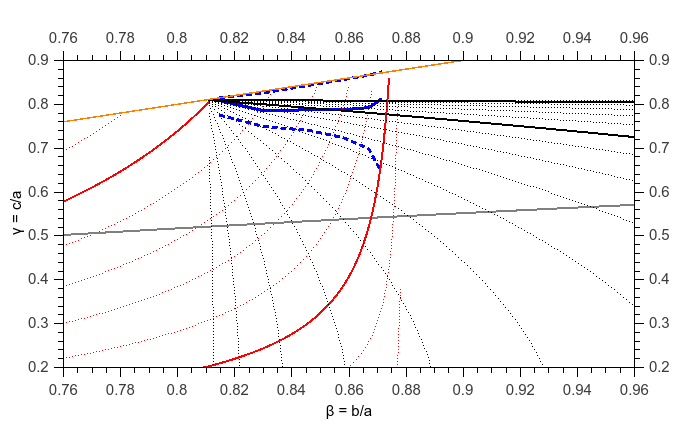}
\caption{Relation between $\beta = b/a$ and $\gamma = c/a$ combining eq.~\ref{eq:gamma_bprime} (black dotted lines -- as in Fig.~\ref{fig:beta_gamma_lellouch_beta_prime}) and eq.~\ref{eq:gamma_deltam} (red dotted lines -- as in Fig.~\ref{fig:beta_gamma_sicardy_delta_m}). Blue curve is the intersection between the black and red lines for each value of $\theta$ and the uncertainties (dotted blue lines -- which depends on the determination of $\beta'$ and $\Delta m$). The orange line is the value for $\beta = \gamma$ while the grey line represents the relation for $\beta$ and $\gamma$ for the Jacobi shape. For $\theta$ between 55 and 80$^\circ$, defined by the black and red full lines, 
there is no intersection between the blue and grey lines, meaning that there is no Jacobi solution for VS2.
\label{fig:beta_gamma_3models}}
\end{center}
\end{figure}

We can plot the two relations between $\gamma$ and $\beta$ for a triaxial body (eq.~\ref{eq:gamma_bprime} is represented in Fig.~\ref{fig:beta_gamma_lellouch_beta_prime} and eq.~\ref{eq:gamma_deltam} in Fig.~\ref{fig:beta_gamma_sicardy_delta_m}) and verify if there exists a set of possible common solutions for every value of $\theta$. Each of the intersections between the two curves for a same $\theta$ are presented in grey line in Fig.~\ref{fig:beta_gamma_3models}. If we also plot the Jacobi relation between $\gamma$ and $\beta$ (blue line in Fig.~\ref{fig:beta_gamma_3models}) we should see a solution that intersects the three curves, which is not the case for VS2. This means that there is a solution for a triaxial shape but this solution is not a Jacobi shape.

In the case of VS2 the best solution for a triaxial shape have values for $\theta = 65 ^{+15}_{-10}$ degrees, with $\beta = 0.846 ^{+0.031}_{-0.028}$ ($b = 265.5 ^{+8.8}_{-9.8}$ km), $\gamma = 0.788 ^{+0.085}_{-0.139}$ ($c = 247.3 ^{+26.6}_{-43.6}$ km). Those values are also given on Table~\ref{tab:solutions_to_VS2}.

In order to have a Jacobi solution, we can explore different values for $\Delta m$, from 0 to the nominal value of 0.141 mag. In fact we are assuming that some of the light curve contribution is due to VS2's shape and some due to albedo variation in the surface. When we try values for $\Delta m$ smaller than 0.141 the lines in Fig.~\ref{fig:beta_gamma_sicardy_delta_m} will move to the right and so more intersections with the lines from Fig.~\ref{fig:beta_gamma_lellouch_beta_prime} will be available, i.e., there will be an intersection between the grey and the blue lines in Fig.~\ref{fig:beta_gamma_3models}. For $\Delta m = 0.015$ mag we find a Jacobi solution with $\beta = 0.908$ ($b = 284.9$ km), $\gamma = 0.553$ ($c = 173.5$ km) and $\theta = 75~^\circ$.


\begin{figure}
\begin{center}
\includegraphics[width=.8\columnwidth]{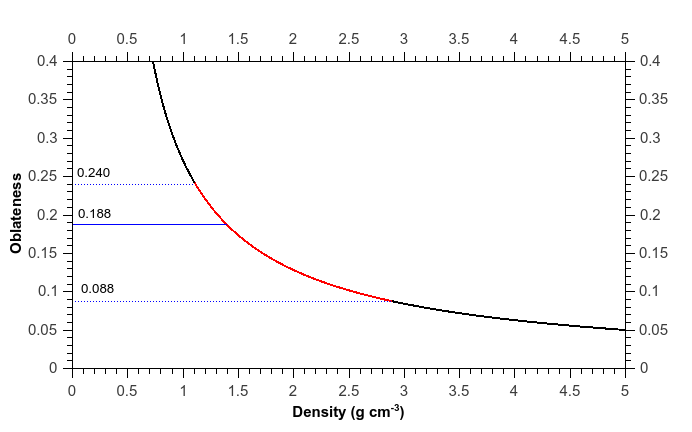}
\caption{Relation between $\epsilon$ and density for a Maclaurin object with rotation period of 7.4175285 hours (black line). The oblateness obtained for VS2 (blue horizontal full and dotted lines) gives limits for the minimum value of the density of 1.4 g cm$^{-3}$. Blue dotted lines represent the values considering the uncertainties in $a$ and $b$. Red line is the interval between the limits.
\label{fig:maclaurin}}
\end{center}
\end{figure}

We can also explore the other extreme and assume that VS2 has an oblate Maclaurin shape (with semi-axis $a=b= 313.8 \pm 7.1 $ km and $254.8 \le c \le 313.8$ km). This assumption automatically imposes that the light curve amplitude variation is only due to albedo features on VS2's surface. Considering the light curve amplitude of 0.141 $\pm$ 0.009 mag -- see Table~\ref{tab:previous_publications} -- the presence of surface irregularities (lumps) and some albedo variegation (spots) on the object with a size approximately of 100 km ($\sim16\%$ of the object equivalent area) is needed. Note that the \textit{Sputnik Planitia} in Pluto is nearly 1000 x 800 km across ($\sim15\%$ of its equivalent area) \citep{Hamilton2016}. Considering the Maclaurin shape and using the rotational period (P = 7.4175285 $\pm$ 0.00001 h -- see Table~\ref{tab:previous_publications}), we can derive VS2's density of $\rho = 1.4 ^{+1.0}_{-0.3}$ g cm$^{-3}$, as show in Fig.~\ref{fig:maclaurin}.


\end{document}